\documentclass[preprint,a4paper,10pt,3p]{elsarticle}

\usepackage{amssymb}
\usepackage{lineno}

\newdefinition{rmk}{Remark}

\usepackage{physics}

\usepackage{xcolor}

\usepackage{psfrag}

\usepackage{algorithm}
\usepackage[end]{algpseudocode}

\usepackage{setspace}

\usepackage{subfigure}

\usepackage{pgfplots}
\usepackage{pgfplotstable}
\pgfplotsset{
tick label style={font=\footnotesize},
label style={font=\small},
axis lines=box,
unbounded coords=jump,
scale only axis,
every axis plot/.append style = {line width = 0.5pt},
ylabel near ticks,
xlabel near ticks,
compat = 1.3
}

\usepackage{tikz}
\usetikzlibrary{patterns}

\journal{Computer Methods in Applied Mechanics and Engineering}

\begin{document}

\begin{frontmatter}


\title{A novel smoothed particle hydrodynamics and finite element coupling scheme for fluid-structure interaction: the sliding boundary particle approach}

\author[1,2]{Sebastian L. Fuchs\corref{cor}}
\ead{fuchs@lnm.mw.tum.de}

\author[1]{Christoph Meier}
\ead{meier@lnm.mw.tum.de}

\author[1]{Wolfgang A. Wall}
\ead{wall@lnm.mw.tum.de}

\author[2,3]{Christian J. Cyron}
\ead{christian.cyron@tuhh.de}

\cortext[cor]{corresponding author}

\address[1]{Institute for Computational Mechanics, Technical University of Munich, Boltzmannstrasse 15, 85748, Garching, Germany}

\address[2]{Institute of Continuum and Materials Mechanics, Hamburg University of Technology, Eissendorfer Str. 42, 21073, Hamburg, Germany}

\address[3]{Institute of Material Systems Modeling, Helmholtz-Zentrum Hereon, Max-Planck-Stra{\ss}e 1, 21502 Geesthacht, Germany}

\begin{abstract}
A novel numerical formulation for solving fluid-structure interaction (FSI) problems is proposed where the fluid field is spatially discretized using smoothed particle hydrodynamics (SPH) and the structural field using the finite element method (FEM). As compared to fully mesh- or grid-based FSI frameworks, due to the Lagrangian nature of SPH this framework can be easily extended to account for more complex fluids consisting of multiple phases and dynamic phase transitions. Moreover, this approach facilitates the handling of large deformations of the fluid domain respectively the fluid-structure interface without additional methodological and computational efforts. In particular, to achieve an accurate representation of interaction forces between fluid particles and structural elements also for strongly curved interface geometries, the novel sliding boundary particle approach is proposed to ensure full support of SPH particles close to the interface. The coupling of the fluid and the structural field is based on a Dirichlet-Neumann partitioned approach, where the fluid field is the Dirichlet partition with prescribed interface displacements and the structural field is the Neumann partition subject to interface forces. To overcome instabilities inherent to weakly coupled schemes an iterative fixed-point coupling scheme is employed. Several numerical examples in form of well-known benchmark tests are considered to validate the accuracy, stability, and robustness of the proposed formulation. Finally, the filling process of a highly flexible thin-walled balloon-like container is studied, representing a model problem close to potential application scenarios of the proposed scheme in the field of biomechanics.
\end{abstract}




\end{frontmatter}

\section{Introduction} \label{sec:intro}

In many applications in science and engineering fluid-structure interaction (FSI) phenomena play an essential role in modeling and simulation, in particular, in some areas of biomechancis, e.g., digestion of food in the human stomach~\cite{Brandstaeter2018, Brandstaeter2019}, referring to the authors target application. Besides the challenge to deal with large deformations of both fluid and structural domain, accurate modeling of fluid flow in biomechanics is even more demanding in the case of complex fluids including, e.g., multiple fluid phases and dynamic phase transitions (e.g. due to chemical reactions). Most current FSI frameworks utilize mesh- or grid-based methods, e.g., the finite element method (FEM), finite difference method (FDM), or finite volume method (FVM), which often require additional methodological and computational effort to capture the aforementioned phenomena. A promising approach to model complex fluids, e.g., the content of gastric lumen in the human stomach~\cite{Brandstaeter2018, Brandstaeter2019}, is the method of smoothed particle hydrodynamics (SPH). SPH is a mesh-free discretization scheme that was originally and independently of one another introduced by Gingold and Monaghan~\cite{Gingold1977} and Lucy~\cite{Lucy1977} in 1977. While initially intended to study astrophysical problems, SPH gained increasing importance in other fields of computational fluid dynamics (CFD) since then. Due to its Lagrangian nature, SPH is very well suited for flow problems involving multiple phases, dynamic phase transitions, as well as complex interface topologies. Especially for many fluid-structure interaction scenarios in biomechanics it would therefore be desirable to discretize the fluid field with SPH whereas the solid field is often easier to handle with finite elements. To this end, a robust and efficient algorithm coupling SPH and FEM for the simulation of fluid-structure interactions is required.

On these grounds, this contribution proposes a novel numerical formulation for solving FSI problems where the fluid field is modeled using SPH and the structural field using FEM. Both sub-fields are coupled following a Dirichlet-Neumann partitioned approach. The fluid field is the Dirichlet partition with prescribed interface displacements or interface velocities, respectively, and the structural field is the Neumann partition subject to interface forces. That means, the interface forces are evaluated by the fluid solver utilizing the current interface displacements and interface velocities that are directly extracted from the structural field. Afterwards, the interface forces are applied to the structural solver enforcing conservation of linear momentum. An iterative fixed-point coupling scheme~\cite{Kuttler2008} is employed to satisfy dynamic equilibrium at the fluid-structure interface with respect to a predefinded convergence criterion. This so-called strong coupling of both sub-fields is crucial to overcome instabilities, e.g., due to the artificial added mass effect, that are known to occur for weakly coupled schemes in FSI~\cite{Causin2005,Forster2007}.

One focus of this work lies on the crucial aspect of the treatment of deformable and strongly curved boundaries of the SPH domain as especially required for many FSI applications. In the literature several different formulations for modeling (rigid) boundaries in SPH are proposed. Among them are penalty-like repulsive force formulations~\cite{Monaghan1994,Monaghan2005a,Monaghan2009}, ghost particle formulations~\cite{Randles1996}, boundary particle methods based on fixed layers of particles resembling rigid walls~\cite{Morris1997,Adami2012}, or semi-analytical methods considering non-vanishing surface integrals due to missing kernel support~\cite{Kulasegaram2004,Ferrand2013,Mayrhofer2015}. For an overview on the advantages and disadvantages of the aforementioned methods the interested reader is refered to the literature, e.g., in~\cite{Liu2010,Ye2019}. In principle, all those methods modeling rigid boundaries in SPH naturally have the potential to serve as a basis also for the treatment of flexible structural boundaries in the context of FSI problems~\cite{Muller2004,Hu2014,Li2015,Fourey2017,Long2017}. However, FSI applications, especially in biomechanics, are characterized by large deformations at the fluid-structure interface including strong curvature and large stretch. This requires a special treatment of boundaries in order to prevent loss of accuracy at the fluid-structure interface. To the best of the authors' knowledge, the existing methods are either missing the required accuracy, computationally expensive, or not capable of modeling deforming interfaces undergoing strong curvature and large stretch. To address this shortcoming of existing approaches, the novel sliding boundary particle approach is proposed. It is based on a transient set of virtual boundary particles regulary arranged around the current projection point of a fluid particle onto the fluid domain boundary. Moreover, a generalized formulation for the extrapolation of field variables from fluid to virtual boundary particles is proposed, which is inspired by the procedure of~\cite{Adami2012}.

The present publication is organized as follows: To begin with, the governing equations for FSI problems are briefly introduced in Section~\ref{sec:goveq}, followed by a detailed presentation of the numerical methods and the computational framework being utilized with a focus on the evaluation of the interface forces and the coupling scheme, cf. Section~\ref{sec:nummeth}. Finally, numerical results obtained with the proposed novel numerical formulation for solving FSI problems are shown in Section~\ref{sec:numex}. For validation purposes, well-known CFD respectively FSI benchmark tests are studied confirming the accuracy and robustness of the proposed formulation. This is followed by an application-motivated academic example examining the filling process of a highly flexible thin-walled container.

\section{Governing equations} \label{sec:goveq}

\begin{figure}[htbp]
\centering
\newcommand*{\scaletext}{1.0}
\newcommand*{\scalefig}{0.75}
\psfrag{os}{\scalebox{\scaletext}{$\Omega^{s}$}}
\psfrag{of}{\scalebox{\scaletext}{$\Omega^{f}$}}
\psfrag{gsd}{\scalebox{\scaletext}{$\Gamma^{s}_{D}$}}
\psfrag{gsn}{\scalebox{\scaletext}{$\Gamma^{s}_{N}$}}
\psfrag{gfd}{\scalebox{\scaletext}{$\Gamma^{f}_{D}$}}
\psfrag{gfn}{\scalebox{\scaletext}{$\Gamma^{f}_{N}$}}
\psfrag{gfs}{\scalebox{\scaletext}{$\Gamma^{fs}$}}
\includegraphics[scale=\scalefig]{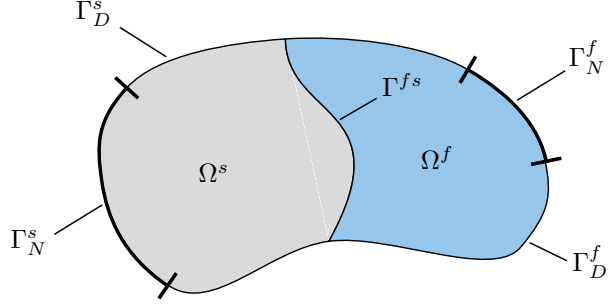}
\caption{Domain~$\Omega$ of a fluid-structure interaction problem consisting of two disjunct sub-domains for the fluid field~$\Omega^{f}$ and the structural field~$\Omega^{s}$ with shared common interface~$\Gamma^{fs}$.}
\label{fig:domain_continuous}
\end{figure}

At all times $t \in \qty[0,T]$ the domain~$\Omega$ of a fluid-structure interaction problem consists of a non-overlapping fluid domain~$\Omega^{f}$ and a structural domain~$\Omega^{s}$ that share a common interface~$\Gamma^{fs}$, with $\Omega = \Omega^{f} \cup \Omega^{s}$ and $\Omega^{f} \cap \Omega^{s} = \Gamma^{fs}$, refer to Figure~\ref{fig:domain_continuous}. This leads to the so-called geometric coupling condition that restricts the fluid and structural domains to perfectly match without any holes or gaps at the fluid-structure interface~$\Gamma^{fs}$. In the following, the (standard) governing equations of the fluid and structural field as well as the respective coupling condition for FSI are briefly given.

% begin figure
% end figure

% begin remark
\begin{rmk}
In the equations~\eqref{eq:fluid_conti}-\eqref{eq:fluid_momentum} governing the fluid field and~\eqref{eq:struct_momentum} governing the structural field all time derivatives follow the motion of material points, i.e., the material derivative reads $\dv{\qty(\cdot)}{t} = \pdv{\qty(\cdot)}{t} + \vectorbold{u} \vdot \grad{\qty(\cdot)}$. Furthermore, $\grad{\qty(\cdot)}$ denotes within the setting of nonlinear continuum mechanics derivatives with respect to spatial coordinates while $\grad_{0}{\qty(\cdot)}$ denotes derivatives with respect to material coordinates.
\end{rmk}
% end remark

%%
\subsection{Fluid field} \label{subsec:goveq_fluid}

The fluid field is governed by the instationary Navier-Stokes equations in the domain~$\Omega^{f}$ in convective form consisting of the mass continuity equation and the momentum equation
\begin{equation} \label{eq:fluid_conti}
\dv{\rho^{f}}{t} = -\rho^{f} \div \vectorbold{u}^{f} \qin \Omega^{f} \, ,
\end{equation}
\begin{equation} \label{eq:fluid_momentum}
\dv{\vectorbold{u}^{f}}{t} = -\frac{1}{\rho^{f}} \grad{p^{f}} + \vectorbold{f}_{\nu} + \vectorbold{b}^{f} \qin \Omega^{f} \, ,
\end{equation}
with viscous force~$\vectorbold{f}_{\nu}$ and body force~$\vectorbold{b}^{f}$ each per unit mass. For a Newtonian fluid the viscous force is $\vectorbold{f}_{\nu} = \nu^{f} \laplacian{\vectorbold{u}^{f}}$ with kinematic viscosity~$\nu^{f}$. The mass continuity equation~\eqref{eq:fluid_conti} and the momentum equation~\eqref{eq:fluid_momentum} represent a system of four equations with the five unknowns, velocity~$\vectorbold{u}^{f}$, density~$\rho^{f}$, and pressure~$p^{f}$. The system of equations is closed with an equation of state $p^{f} = p^{f}\qty(\rho^{f})$ relating fluid density~$\rho^{f}$ and pressure~$p^{f}$, cf. Section~\ref{subsec:nummeth_sph_eos}. The Navier-Stokes equations~\eqref{eq:fluid_conti} and~\eqref{eq:fluid_momentum} are subject to the following initial conditions
\begin{equation}
\rho^{f} = \rho^{f}_{0} \qand \vectorbold{u}^{f} = \vectorbold{u}^{f}_{0} \qin \Omega^{f} \qq{at} t = 0
\end{equation}
with initial density $\rho^{f}_{0}$ and initial velocity~$\vectorbold{u}^{f}_{0}$. In addition, Dirichlet and Neumann boundary conditions are applied on the fluid boundary~$\Gamma^{f} = \partial\Omega^{f} \setminus \Gamma^{fs}$
\begin{equation}
\vectorbold{u}^{f} = \vectorbold{\hat{u}}^{f} \qq{on} \Gamma^{f}_{D} \qand
\vectorbold{t}^{f} = \vectorbold{\hat{t}}^{f} \qq{on} \Gamma^{f}_{N} \, ,
\end{equation}
with prescribed boundary velocity~$\vectorbold{\hat{u}}^{f}$ and boundary traction~$\vectorbold{\hat{t}}^{f}$, where $\Gamma^{f} = \Gamma^{f}_{D} \cup \Gamma^{f}_{N}$ and $\Gamma^{f}_{D} \cap \Gamma^{f}_{N} = \emptyset$.

\subsection{Structural field} \label{subsec:goveq_struct}

Considering the regime of finite deformations, the structural field is governed by the balance of linear momentum in the following local material form
\begin{equation} \label{eq:struct_momentum}
\rho^{s}_{0} \dv[2]{\vectorbold{d}^{s}}{t} = \grad_{0} \vdot \qty(\vectorbold{F} \vectorbold{S}) + \rho^{s}_{0} \vectorbold{b}^{s}_{0} \qin \Omega^{s}
\end{equation}
with the material forms of density~$\rho^{s}_{0}$ and body force~$\vectorbold{b}^{s}_{0}$, and the structural displacement~$\vectorbold{d}^{s}$ as primary unknowns. The deformation of the structure is described by the deformation gradient $\vectorbold{F} = \grad_{0} {\vectorbold{d}^{s}}$ defining the Green-Lagrange strains $\vectorbold{E} = \frac{1}{2} \qty(\vectorbold{F}^{T} \vectorbold{F} - \vectorbold{I})$. For simplicity, and as applicable and most often used in biomechanical problems, the second Piola-Kirchhoff stresses $\vectorbold{S}$ are chosen to follow from a constitutive relation of the form $\vectorbold{S} = \pdv*{\Psi}{\vectorbold{E}}$ based on a hyperelastic strain energy function $\Psi = \Psi\qty(\vectorbold{E})$. The partial differential equation~\eqref{eq:struct_momentum} is subject to initial conditions for the structural displacement and velocity
\begin{equation}
\vectorbold{d}^{s} = \vectorbold{d}^{s}_{0} \qq{and} \dv{\vectorbold{d}^{s}}{t} = \dv{\vectorbold{d}^{s}_0}{t} \qin \Omega^{s} \qq{at} t = 0 \, .
\end{equation}
On the structural boundary~$\Gamma^{s} = \partial\Omega^{s} \setminus \Gamma^{fs}$, Dirichlet and Neumann boundary conditions are prescribed
\begin{equation}
\vectorbold{d}^{s} = \vectorbold{\hat{d}}^{s} \qq{on} \Gamma^{s}_{D} \qand
\qty(\vectorbold{F} \vectorbold{S}) \vdot \vectorbold{N} = \vectorbold{\hat{t}}^{s}_{0}  \qq{on} \Gamma^{s}_{N} \, ,
\end{equation}
with prescribed boundary displacement~$\vectorbold{\hat{d}}^{s}$, boundary traction~$\vectorbold{\hat{t}}^{s}_{0}$, and outward pointing unit normal vector~$\vectorbold{N}$ on~$\Gamma^{s}$ in material description, where $\Gamma^{s} = \Gamma^{s}_{D} \cup \Gamma^{s}_{N}$ and $\Gamma^{s}_{D} \cap \Gamma^{s}_{N} = \emptyset$.

\subsection{Coupling conditions} \label{subsec:goveq_coup}

A geometric coupling condition results from restricting both the fluid and structural domain to match at the fluid-structure interface~$\Gamma^{fs}$ as already described in the beginning of this section. In addition, the so-called kinematic coupling condition (or no-slip boundary condition) enforces a continuous fluid and structural velocity at the interface~$\Gamma^{fs}$. Consequently, these two conditions can be expressed as
\begin{equation} \label{eq:coup_geom_kinem}
\vectorbold{r}^{f} = \vectorbold{r}^{s} \qand \vectorbold{u}^{f} = \dv{\vectorbold{d}^{s}}{t} \qq{on} \Gamma^{fs} \, ,
\end{equation}
with the current position~$\vectorbold{r}^{f}$ respectively~$\vectorbold{r}^{s}$ of the fluid and structural field. Finally, the dynamic coupling condition ensures equilibrium of fluid and structural traction across the interface~$\Gamma^{fs}$
\begin{equation} \label{eq:coup_dyn}
\vectorbold{t}^{f} = \vectorbold{t}^{s} \qq{on} \Gamma^{fs} \, .
\end{equation}

\section{Numerical methods and computational framework} \label{sec:nummeth}

\begin{figure}[htbp]
\centering
\newcommand*{\scaletext}{1.0}
\newcommand*{\scalefig}{0.75}
\psfrag{os}{\scalebox{\scaletext}{$\Omega^{s}$}}
\psfrag{of}{\scalebox{\scaletext}{$\Omega^{f}$}}
\psfrag{gfs}{\scalebox{\scaletext}{$\Gamma^{fs}$}}
\psfrag{str}{\scalebox{\scaletext}{structural element}}
\psfrag{int}{\scalebox{\scaletext}{interface element}}
\psfrag{part}{\scalebox{\scaletext}{fluid particle}}
\psfrag{svs}{\scalebox{\scaletext}{structural solver (FEM)}}
\psfrag{svf}{\scalebox{\scaletext}{fluid solver (SPH)}}
\psfrag{id}{\scalebox{\scaletext}{$\vectorbold{d}^{fs}$}}
\psfrag{if}{\scalebox{\scaletext}{$\vectorbold{f}^{fs}$}}
\includegraphics[scale=\scalefig]{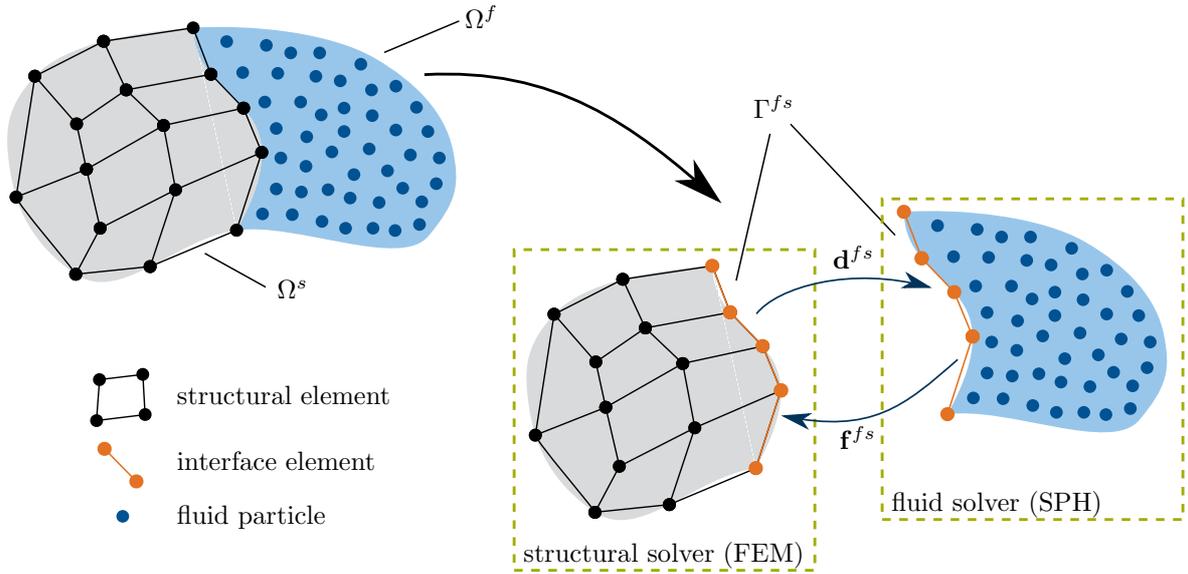}
\caption{Discretized domain~$\Omega$ of a fluid-structure interaction problem with structural mesh and fluid particles (left) and separated sub-domains as seen by the fluid solver (SPH) and the structural solver (FEM) each with interface mesh for interaction handling and exchange of interface displacements~$\vectorbold{d}^{fs}$ and interface forces~$\vectorbold{f}^{fs}$ in the sense of a Dirichlet-Neumann partitioned coupling approach (right).}
\label{fig:domain_discretized}
\end{figure}

The purpose of this section is to present the methods for discretization and numerical solution of the fluid-structure interaction problem as described in Section~\ref{sec:goveq}. The discretization of the fluid field is based on smoothed particle hydrodynamics while the discretization of the structural field is based on the finite element method, as illustrated in Figure~\ref{fig:domain_discretized} (left).

% begin figure
% end figure

While in Sections~\ref{subsec:nummeth_sph} and \ref{subsec:nummeth_fem} the basics of these two methods are recapitulated, the focus of this publication is set on the specific evaluation of interaction forces, cf. Section~\ref{subsec:nummeth_inter}, introducing the sliding boundary particle approach, and the employed coupling algorithm, cf. Section~\ref{subsec:nummeth_coupappr}, in terms of underlying methods. The presented computational framework is implemented in the in-house parallel multiphysics research code BACI (Bavarian Advanced Computational Initiative) \cite{Baci}.

\subsection{Discretization of fluid field via smoothed particle hydrodynamics} \label{subsec:nummeth_sph}

The fluid field governed by the instationary Navier-Stokes equations~\eqref{eq:fluid_conti} and~\eqref{eq:fluid_momentum} is solved using smoothed particle hydrodynamics following a weakly compressible approach~\cite{Monaghan2005a,Liu2010,Price2012}. For modeling fluid flow using SPH, several different formulations each with its own characteristics and benefits can be derived as reflected by the vast amount of literature. The aim of this section is to give a brief introduction into the basics of SPH and an overview of the formulation applied throughout this work. Note that the contribution resulting from the coupling condition of the fluid and structural field at the interface~$\Gamma^{fs}$, refer to Section~\ref{subsec:goveq_coup}, is omitted in this section and described in detail in Section~\ref{subsec:nummeth_inter}. For ease of notation, in the following the index~$\qty(\cdot)^{f}$ denoting fluid quantities, as used in Section~\ref{sec:goveq}, is dropped.

\subsubsection{Approximation of field quantities via smoothing kernel} \label{subsec:nummeth_sph_kernel}

The fundamental concept of SPH is based on the approximation of a field quantity~$f$ via a smoothing operation and on the discretization of the domain~$\Omega$ with discretization points, so-called particles. To begin with, a field quantity~$f$ on a domain~$\Omega$ can be expressed exactly in integral form as
\begin{equation}
f\qty(\vectorbold{r}) = \int_{\Omega} f\qty(\vectorbold{r}') \delta\qty(\qty| \vectorbold{r} - \vectorbold{r}' |) \dd{\vectorbold{r}'}
\end{equation}
making use of the Dirac delta function~$\delta\qty(r)$. Replacing the latter by a so-called smoothing kernel~$W\qty(r,h)$, that fulfills certain required properties, cf. Remark~\ref{rmk:sph_kernel_properties} and \cite{Liu2010}, leads to an approximation of the field quantity~$f$ in smoothed integral form
\begin{equation}
f\qty(\vectorbold{r}) \approx \int_{\Omega} f\qty(\vectorbold{r}') W\qty(\qty| \vectorbold{r} - \vectorbold{r}' |, h) \dd{\vectorbold{r}'}
\end{equation}
while committing a smoothing error.

% begin remark
\begin{rmk} \label{rmk:sph_kernel_properties}
The smoothing kernel~$W\qty(r,h)$ is a monotonically decreasing, smooth function that depends on a distance~$r$ and a smoothing length~$h$. The smoothing length~$h$ together with a scaling factor~$\kappa$ define the support radius of the smoothing kernel~$r_{c} = \kappa h$. Compact support, i.e., $W\qty(r, h) = 0$ for $r > r_{c}$, as well as positivity, i.e., $W\qty(r, h) \geq 0$ for $r \leq r_{c}$, are typical properties of standard smoothing kernels~$W\qty(r,h)$. In addition, the normalization property requires that $\int_{\Omega} W\qty(\qty| \vectorbold{r} - \vectorbold{r}' |, h) \dd{\vectorbold{r}'} = 1$. The Dirac delta function property $\lim_{h \rightarrow 0}{ W\qty(r, h) } = \delta\qty(r)$ ensures an exact representation of a field quantity~$f$ in the limit $h \rightarrow 0$.
\end{rmk}
% end remark

In a next step, the computational domain~$\Omega$ is filled with discretization points or so-called particles~$j$, each occupying a volume~$V_{j}$. Thus, the smoothed integral form of quantity~$f$ reduces in discretized form to a summation of contributions from all particles~$j$ in the domain~$\Omega$, cf. Remark~\ref{rmk:sph_particles_in_summation},
\begin{equation} \label{eq:sph_approximation}
f\qty(\vectorbold{r}) \approx \sum_{j} V_{j} f\qty(\vectorbold{r}_{j}) W\qty(\qty| \vectorbold{r} - \vectorbold{r}_{j} |, h)
\end{equation}
adding a discretization error~\cite{Quinlan2006}. A straightforward approach in SPH to determine the gradient of a quantity~$f$ follows directly by differentiation of equation \eqref{eq:sph_approximation} resulting in
\begin{equation}
\grad{f\qty(\vectorbold{r})} \approx \sum_{j} V_{j} f\qty(\vectorbold{r}_{j}) \grad{W\qty(\qty| \vectorbold{r} - \vectorbold{r}_{j} |, h)} \, .
\end{equation}
Note that this (simple) variant for an approximation of the gradient shows some particular disadvantages, hence, more advanced approximations for gradients are given in the literature \cite{Monaghan2005a} and also applied in this work \cite{Adami2012,Adami2013}, cf. Section~\ref{subsec:nummeth_sph_momentum}.

% begin remark
\begin{rmk} \label{rmk:sph_particles_in_summation}
In general, contributions from all particles in the domain~$\Omega$ are considered in the SPH approximation of a field quantity~$f$, cf. equation \eqref{eq:sph_approximation}. However, note that in practice due to the compact support of the smoothing kernel~$W$ only neighboring particles within the support radius~$r_{c}$ need to be considered. This property is very beneficial as it reduces the computational effort of the method.
\end{rmk}
% end remark

Applying the concept of SPH reduces the partial differential equations~\eqref{eq:fluid_conti} and~\eqref{eq:fluid_momentum} to ordinary differential equations that are solved, i.e., evaluated and integrated in time, for all particles in the domain~$\Omega$ (cf. Sections \ref{subsec:nummeth_sph_momentum} and \ref{subsec:nummeth_sph_timint}). The transient positions of particles are advected with the fluid velocity resembling the Lagrangian nature of the method. As a result, all fluid quantities are evaluated at and associated with particle positions, meaning each particle carries its corresponding fluid quantities.

Finally, in a post-processing step the continuous field quantity $f$ is recovered from the discrete fluid quantities carried by each particle in the domain based on approximation \eqref{eq:sph_approximation} and the commonly known Shepard filter
\begin{equation} \label{eq:sph_postprocessing}
\hat{f}\qty(\vectorbold{r}) \approx \frac{ \sum_{j} V_{j} f\qty(\vectorbold{r}_{j}) W\qty(\qty| \vectorbold{r} - \vectorbold{r}_{j} |, h) }{ \sum_{j} V_{j} W\qty(\qty| \vectorbold{r} - \vectorbold{r}_{j} |, h) } \, .
\end{equation}
Note that the denominator typically takes on values close to one inside the fluid domain and is mainly relevant for boundary regions with reduced support due to a lack of neighboring particles.

% begin remark
\begin{rmk}
In the following, a quantity~$f$ evaluated for particle~$i$ at position~$\vectorbold{r}_{i}$ is written as~$f_{i} = f\qty(\vectorbold{r}_{i})$. In addition, the short notation $W_{ij} = W\qty(r_{ij}, h)$ denotes the smoothing kernel~$W$ evaluated for particle~$i$ at position~$\vectorbold{r}_{i}$ with neighboring particle~$j$ at position~$\vectorbold{r}_{j}$, where $r_{ij} = \qty|\vectorbold{r}_{ij}| = \qty|\vectorbold{r}_{i} - \vectorbold{r}_{j}|$ is the absolute distance between particles~$i$ and~$j$. Similarly, the derivative of the smoothing kernel~$W$ with respect to the absolute distance~$r_{ij}$ is denoted by $\pdv*{W}{r_{ij}} = \pdv*{W\qty(r_{ij}, h)}{r_{ij}}$.
\end{rmk}
% end remark

% begin remark
\begin{rmk}
Herein, the smoothing of fluid quantities is carried out using a quintic spline smoothing kernel~$W\qty(r, h)$ as defined in \cite{Morris1997} with smoothing length~$h$ and compact support of the smoothing kernel with support radius~$r_{c} = \kappa h$ and scaling factor~$\kappa = 3$.
\end{rmk}
% end remark

%%
\subsubsection{Initial particle spacing} \label{subsec:nummeth_sph_spacing}

Within this contribution, the fluid domain is initially filled with particles located on a regular grid with particle spacing~$\Delta{}x$, thus in a $d$-dimensional space each particle initially occupies an effective volume of~$\qty(\Delta{}x)^{d}$. The mass of a particle~$i$ is then set using the reference density according to $m_{i} = \rho_{0} \qty(\Delta{}x)^{d}$ and remains constant throughout the simulation. In general, the initial particle spacing~$\Delta{}x$ can be freely chosen, however, within this work the initial particle spacing~$\Delta{}x$ is set equal to the smoothing length~$h = \flatfrac{r_{c}}{\kappa}$.

\subsubsection{Density summation} \label{subsec:nummeth_sph_summation}

The density of a particle~$i$ is determined via summation of the respective smoothing kernel contributions of all neighboring particles~$j$ within the support radius~$r_{c}$
\begin{equation} \label{eq:sph_densum}
\rho_{i} = m_{i} \sum_{j} W_{ij} \, .
\end{equation}
This approach is typically denoted as density summation and results in an exact conservation of mass in the fluid domain, which can be shown in a straightforward manner considering the commonly applied normalization of the smoothing kernel to unity. It shall be noted that the density field may alternatively be obtained by discretization and integration of the mass continuity equation~\eqref{eq:fluid_conti}~\cite{Liu2010}.

\subsubsection{Momentum equation} \label{subsec:nummeth_sph_momentum}

The momentum equation~\eqref{eq:fluid_momentum} is discretized following~\cite{Adami2012, Adami2013} including a transport velocity formulation to suppress the problem of tensile instability. It will be briefly recapitulated in the following. The transport velocity formulation relies on a constant background pressure~$p_{b}$ that is applied to all particles and results in a contribution to the particle accelerations for in general disordered particle distributions. However, these additional acceleration contributions vanish for particle distributions fulfilling the partition of unity, thus fostering these desirable configurations. For the sake of brevity, the definition of the modified advection velocity and the additional terms in the momentum equation from the aforementioned transport velocity formulation are not discussed in the following and the reader is kindly referred to the original publication~\cite{Adami2013}. Altogether, the acceleration $\vectorbold{a}_{i} = \dv*{\vectorbold{u}_{i}}{t}$ of a particle~$i$ results from summation of all acceleration contributions due to interaction with neighboring particles~$j$ and a body force as
\begin{equation} \label{eq:sph_momentum}
\vectorbold{a}_{i} = \frac{1}{m_{i}} \sum_{j} \qty(V_{i}^{2}+V_{j}^{2}) \qty[ - \tilde{p}_{ij} \pdv{W}{r_{ij}} \vectorbold{e}_{ij} + \tilde{\eta}_{ij} \frac{\vectorbold{u}_{ij}}{r_{ij}} \pdv{W}{r_{ij}} ] + \vectorbold{b}_{i} \, ,
\end{equation}
with volume~$V_{i} = m_{i}/\rho_{i}$ of particle~$i$, unit vector $\vectorbold{e}_{ij} = \flatfrac{\vectorbold{r}_{i} - \vectorbold{r}_{j}}{\qty|\vectorbold{r}_{i} - \vectorbold{r}_{j}|} = \flatfrac{\vectorbold{r}_{ij}}{r_{ij}}$ pointing from particle~$j$ to particle~$i$, relative velocity $\vectorbold{u}_{ij} = \vectorbold{u}_{i}-\vectorbold{u}_{j}$, density-weighted inter-particle averaged pressure
\begin{equation} \label{eq:sph_mom_wght_press}
\tilde{p}_{ij} = \frac{\rho_{j}p_{i}+\rho_{i}p_{j}}{\rho_{i} + \rho_{j}} \, ,
\end{equation}
and inter-particle averaged dynamic viscosity
\begin{equation} \label{eq:sph_mom_wght_visc}
\tilde{\eta}_{ij} = \frac{2\eta_{i}\eta_{j}}{\eta_{i}+\eta_{j}} \, .
\end{equation}
In the following the acceleration contribution of a neighboring particle~$j$ to particle~$i$ is, for ease of notation, denoted as~$\vectorbold{a}_{ij}$, where $\vectorbold{a}_{i} = \sum_{j} \vectorbold{a}_{ij} + \vectorbold{b}_{i}$. Note that the above given momentum formulation, cf. equation~\eqref{eq:sph_momentum}, exactly conserves linear momentum due to pairwise anti-symmetric particle forces
\begin{equation} \label{eq:sph_conservlinmom}
m_{i} \vectorbold{a}_{ij} = - m_{j} \vectorbold{a}_{ji} \, ,
\end{equation}
which can easily be verified by using the property $\pdv*{W}{r_{ij}} = \pdv*{W}{r_{ji}}$ of the smoothing kernel.

\subsubsection{Equation of state} \label{subsec:nummeth_sph_eos}

Following a weakly compressible approach, density~$\rho_{i}$ and pressure~$p_{i}$ of a particle~$i$ are linked via the equation of state
\begin{equation} \label{eq:sph_eos}
p_{i}\qty(\rho_{i}) = c^{2} \qty(\rho_{i} - \rho_{0}) = p_{0} \qty(\frac{\rho_{i}}{\rho_{0}} - 1)
\end{equation}
with reference density~$\rho_{0}$, reference pressure~$p_{0} = \rho_{0} c^{2}$ and artificial speed of sound~$c$. Note that this commonly applied approach only represents deviations from the reference pressure, i.e., $p_{i}\qty(\rho_{0}) = 0$, and not the total pressure. Thus, free boundaries can be modeled by setting~$p = 0$ (see also Section~\ref{subsec:nummeth_sph_bdrycond} below). To limit density fluctuations to an acceptable level, while still avoiding too severe time step restrictions, strategies are discussed in~\cite{Morris1997} on how to determine an appropriate value of the artificial speed of sound.

\subsubsection{Boundary conditions} \label{subsec:nummeth_sph_bdrycond}
\paragraph{Rigid wall boundary conditions}

Following the approach of~\cite{Adami2012}, rigid wall boundary conditions are modeled using fixed boundary particles with quantities extrapolated from the fluid field based on a local force balance. For more details the interested reader is referred to the aforementioned literature. In the numerical examples in Section~\ref{sec:numex} the channel walls are modeled using rigid wall boundary conditions.

\paragraph{Inflow and outflow boundary conditions}

Open boundaries are modeled similar to~\cite{Lastiwka2009} via defined inflow and outflow zones occupying so-called inflow respectively outflow particles. Thereby, full support of the interior fluid particles is maintained for density summation~\eqref{eq:sph_densum} and evaluation of the momentum equation~\eqref{eq:sph_momentum} when considering contributions from neighboring inflow and outflow particles. At the inflow, i.e., the Dirichlet boundary, the desired inflow velocity is prescribed directly to all inflow particles, while the pressure field is extrapolated from the interior fluid particles~$i$ to the inflow particles~$k$ following
\begin{equation}
p_{k} = \frac{\sum_{i} V_{i} p_{i} W_{ki}}{\sum_{i} V_{i} W_{ki}} \, .
\end{equation}
At the outflow, i.e., the Neumann boundary, a zero pressure field is prescribed to all outflow particles. The density field of both inflow and outflow particles is determined from the pressure field with the equation of state~\eqref{eq:sph_eos}. Finally, to determine consistent velocities of the outflow particles, the momentum equation~\eqref{eq:sph_momentum} is evaluated for outflow particles considering interactions with neighboring fluid particles, boundary particles, and outflow particles.

\paragraph{Periodic boundary conditions}

Imposing a periodic boundary condition in a specific spatial direction allows for particle interaction evaluation across opposite domain borders. Moreover, particles leaving the domain on one side are re-injecting on the opposite side. Periodic boundary conditions are commonly applied in SPH modeling of channel or shear flow.

\subsubsection{Time integration scheme} \label{subsec:nummeth_sph_timint}

The momentum equation~\eqref{eq:sph_momentum} is integrated in time applying an explicit velocity-Verlet time integration scheme in kick-drift-kick form, also denoted as leapfrog scheme, as proposed by Monaghan~\cite{Monaghan2005a}. In the absence of dissipative effects, the velocity-Verlet scheme is of second order accuracy and reversible in time~\cite{Monaghan2005a}.

In a first kick-step the particle accelerations $\vectorbold{a}_{i}^{n} = \qty(\dv*{\vectorbold{u}_{i}}{t})^{n}$ determined in the previous time step~$n$ are used to compute intermediate particle velocities at~$n+1/2$
\begin{equation}
\vectorbold{u}_{i}^{n+1/2} = \vectorbold{u}_{i}^{n} + \frac{\Delta{}t}{2} \, \vectorbold{a}_{i}^{n} \, ,
\end{equation}
where~$\Delta{}t$ is the time step size, before the particle positions at~$n+1$ are updated in a drift-step
\begin{equation}
\vectorbold{r}_{i}^{n+1} = \vectorbold{r}_{i}^{n} + \Delta{}t\vectorbold{u}_{i}^{n+1/2} \, .
\end{equation}
Using the particle positions~$\vectorbold{r}_{i}^{n+1}$ and intermediate velocities~$\vectorbold{u}_{i}^{n+1/2}$, the particle densities~$\rho_{i}^{n+1}$ and accelerations~$\vectorbold{a}_{i}^{n+1}$ are updated following equations~\eqref{eq:sph_densum} and~\eqref{eq:sph_momentum}. In a final kick-step the particle velocities at~$n+1$ are determined
\begin{equation}
\vectorbold{u}_{i}^{n+1} = \vectorbold{u}_{i}^{n+1/2} + \frac{\Delta{}t}{2} \, \vectorbold{a}_{i}^{n+1} \, .
\end{equation}
To maintain stability of the time integration scheme, the time step size~$\Delta{}t$ is restricted by the Courant-Friedrichs-Lewy (CFL) condition, the viscous condition, and the body force condition, refer to~\cite{Morris1997, Adami2013} for more details,
\begin{equation} \label{eq:sph_timestepcond}
\Delta{}t \leq \min\qty{ 0.25\frac{h}{c+\qty|\vectorbold{u}_{max}|}, \quad 0.125\frac{h^{2}}{\nu}, \quad 0.25\sqrt{\frac{h}{\qty|\vectorbold{b}_{max}|}} } \, ,
\end{equation}
with maximum fluid velocity~$\vectorbold{u}_{max}$ and maximum body force~$\vectorbold{b}_{max}$.

\subsection{Discretization of structural field via the finite element method} \label{subsec:nummeth_fem}

The discretization of the structural field, governed by the strong form of the balance of linear momentum~\eqref{eq:struct_momentum}, is based on the finite element method. Since it is not the focus of this work, the basics of the FEM are presented here only very briefly. For further informations the reader is referred to, e.g.,~\cite{Belytschko2013,Zienkiewicz2014}.

Applying the method of weighted residuals, in the following interpreted as principle of virtual work, the weak form of the initial boundary value problem for the structural field is obtained as
\begin{equation} \label{eq:fem_weak}
\delta\mathcal{W}^{s} = \qty(\delta\vectorbold{d}^{s}, \, \rho^{s}_{0} \dv[2]{\vectorbold{d}^{s}}{t})_{\Omega^{s}} + \qty\Big(\grad_{0}{\delta\vectorbold{d}^{s}}, \, \vectorbold{F}\vectorbold{S})_{\Omega^{s}} - \qty\Big(\delta\vectorbold{d}^{s}, \, \rho^{s}_{0}\vectorbold{b}^{s}_{0})_{\Omega^{s}} - \qty\Big(\delta\vectorbold{d}^{s}, \, \vectorbold{\hat{t}}^{s}_{0})_{\Gamma^{s}_{N}} = 0
\end{equation}
with the variation~$\delta\vectorbold{d}^{s}$ of the primary unknown structural displacement~$\vectorbold{d}^{s}$. Herein, the contribution to the weak form resulting from the coupling condition of the fluid and structural field at the interface~$\Gamma^{fs}$ (cf. Section~\ref{subsec:goveq_coup}) is omitted and instead treated in Section~\ref{subsec:nummeth_inter}.

By introducing the trial space $\mathcal{V} = \qty\big{\vectorbold{d}^{s} \, | \, \vectorbold{d}^{s} \in \mathcal{H}^{1}, \, \vectorbold{d}^{s} = \vectorbold{\hat{d}}^{s} \, \text{on} \, \Gamma^{s}_{D}}$ as well as the test space $\mathcal{W} = \qty\big{\delta\vectorbold{d}^{s} \, | \, \delta\vectorbold{d}^{s} \in \mathcal{H}^{1}, \, \delta\vectorbold{d}^{s} = 0 \, \text{on} \, \Gamma^{s}_{D}}$, where~$\mathcal{H}^{1}$ denotes the Sobolev space of functions with square-integrable first derivatives, the weak form~\eqref{eq:fem_weak} is equivalent to the strong form of the balance of linear momentum~\eqref{eq:struct_momentum}.

The computational domain of the structural field~$\Omega^{s}$ is sub-divided into non-overlapping finite elements with nodes $i$. Hence, the structural displacement field~$\vectorbold{d}^{s}$ is discretized introducing nodal displacements~$\vectorbold{d}^{s}_{i}$ of nodes~$i$. The displacement field is approximated via
\begin{equation}
\vectorbold{d}^{s}\qty(\vectorbold{r}) \approx \sum_{j} N^{e}_{j}\qty(\vectorbold{r}) \vectorbold{d}^{s}_{j}
\end{equation}
using the Lagrange polynomials~$N^{e}_{j}$ with compact support inside element~$e$. Within a Bubnov-Galerkin approach, the same Lagrange polynomials for trial and test functions are employed. Following the iso-parametric concept, the parameter coordinates~$\vectorbold*{\xi}$ used for the definition of the shape functions within a standard element geometry are mapped onto the physical coordinates applying the same shape functions also used for the displacement interpolation. Specifically, in the numerical examples in Section~\ref{sec:numex} finite elements based on first-order interpolation are employed.

Subsequently, the semi-discrete form is discretized in time applying a generalized-alpha time integration scheme. The resulting system of nonlinear equations in residual form is finally solved for the nodal structural displacements using a Newton-Raphson method.

\subsection{SPH-FE interaction: a novel sliding boundary particle approach} \label{subsec:nummeth_inter}

\begin{figure}[htbp]
\centering
\newcommand*{\scaletext}{1.0}
\newcommand*{\scalefig}{0.75}
\psfrag{fp}{\scalebox{\scaletext}{fluid particle}}
\psfrag{bp}{\scalebox{\scaletext}{boundary particle}}
\includegraphics[scale=\scalefig]{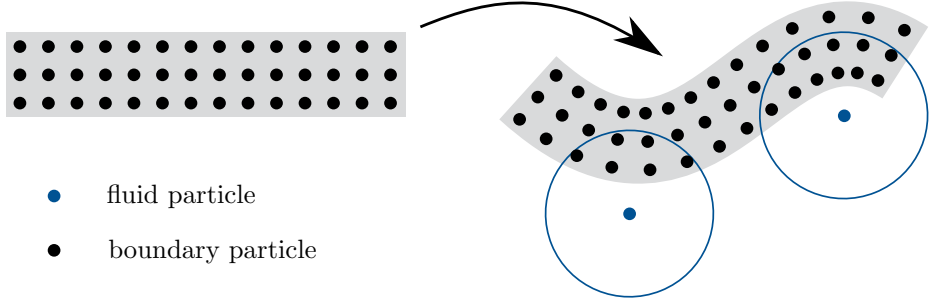}
\caption{A structural domain initially discretized by a regular and equidistant set of boundary particles fixed to material points of the structure (left) undergoing strong curvature and large stretch (right). Clearly, the support of the smoothing kernel of fluid particles close to regions of large structural displacements is disturbed.}
\label{fig:motivation_deforming}
\end{figure}

In this section, a novel sliding boundary particle approach for the application in a fluid-structure interaction framework coupling SPH and FEM is proposed. In contrast to existing methods modeling boundaries in SPH, e.g., boundary particle methods, cf. Figure~\ref{fig:motivation_deforming}, the proposed method can handle also deforming interfaces undergoing strong curvature and large stretch, as typical for some FSI applications especially in biomechanics, while keeping the computational costs at a reasonable level. The following is mainly concerned with the evaluation of the interface force~$\vectorbold{f}^{fs}$ at the fluid-structure interface~$\Gamma^{fs}$. The coupling of fluid and structural field following a Dirichlet-Neumann partitioned approach is subsequently described in Section~\ref{subsec:nummeth_coupappr}.

% begin figure
% end figure

%%
\subsubsection{Conforming interface mesh} \label{subsec:nummeth_inter_interfacemesh}

Introducing an interface mesh on the fluid-structure interface~$\Gamma^{fs}$ allows for exchange of interface displacement~$\vectorbold{d}^{fs}$ and interface force~$\vectorbold{f}^{fs}$ between the fluid and the structural field, cf. Figure~\ref{fig:domain_discretized}, while keeping the fluid and structural solvers separated. For convenience, the interface mesh, which is purely introduced as one possibility to facilitate the displacement and load transfer between the solvers, can be chosen as an extraction or clone of the structural mesh at the fluid-structure interface~$\Gamma^{fs}$. But the proposed approach also works for non-matching meshes. For the interface mesh, again the iso-parametric concept is employed to describe the standard element geometry of interface elements~$e$ via parameter coordinates~$\vectorbold*{\xi}$ and Lagrange polynomials~$N^{e}_{j}$ of corresponding nodes~$j$. Note that for interface elements the parameter coordinates~$\vectorbold*{\xi}$ are of one dimension lower compared to structural elements. Deduced from the geometric coupling condition~\eqref{eq:coup_geom_kinem} the current interface position is in the following depicted by~$\vectorbold{r}^{fs}$.

In case of a conforming mesh, the transfer of quantities between interface and structure is straightforward and is, hence, just briefly sketched here. Both, interface position~$\vectorbold{r}^{fs}$ and interface displacement~$\vectorbold{d}^{fs}$ can be extracted directly from the respective structural position~$\vectorbold{r}^{s}$ and structural displacement~$\vectorbold{d}^{s}$. Similarly, the interface force~$\vectorbold{f}^{fs}$ can be added directly to the respective structural force~$\vectorbold{f}^{s}$. In case non-matching interfaces are preferred or needed, e.g., because of special resolution demands of the two involved physical fields, the transfer of quantities between interface and structure could simply be done via a Mortar technique~\cite{Kloeppel2011}. In comparison to the interface structure transfer, the transfer of quantities to the fluid field is more elaborate and will be covered in the following subsections.

\subsubsection{Detection of closest projection point} \label{subsec:nummeth_inter_closestprojectionpoint}

\begin{figure}[htbp]
\centering
\newcommand*{\scaletext}{1.0}
\newcommand*{\scalefig}{1.0}
\psfrag{gfs}{\scalebox{\scaletext}{$\Gamma^{fs}$}}
\psfrag{r_c}{\scalebox{\scaletext}{support radius~$r_c$}}
\psfrag{fp}{\scalebox{\scaletext}{fluid particle~$i$}}
\psfrag{cp}{\scalebox{\scaletext}{closest projection point~$c^{e}_{i}$}}
\psfrag{vp}{\scalebox{\scaletext}{virtual boundary particle~$k^{e}_{i}$}}
\psfrag{ife}{\scalebox{\scaletext}{interface element~$e$}}
\psfrag{r_ci}{\scalebox{\scaletext}{$\vectorbold{r}_{c^{e}_{i}} - \vectorbold{r}_{i}$}}
\psfrag{r_ki}{\scalebox{\scaletext}{$\vectorbold{r}_{k^{e}_{i}} - \vectorbold{r}_{i}$}}
\psfrag{dx}{\scalebox{\scaletext}{$\Delta{}x$}}
\psfrag{dx2}{\scalebox{\scaletext}{$\flatfrac{\Delta{}x}{2}$}}
\includegraphics[scale=\scalefig]{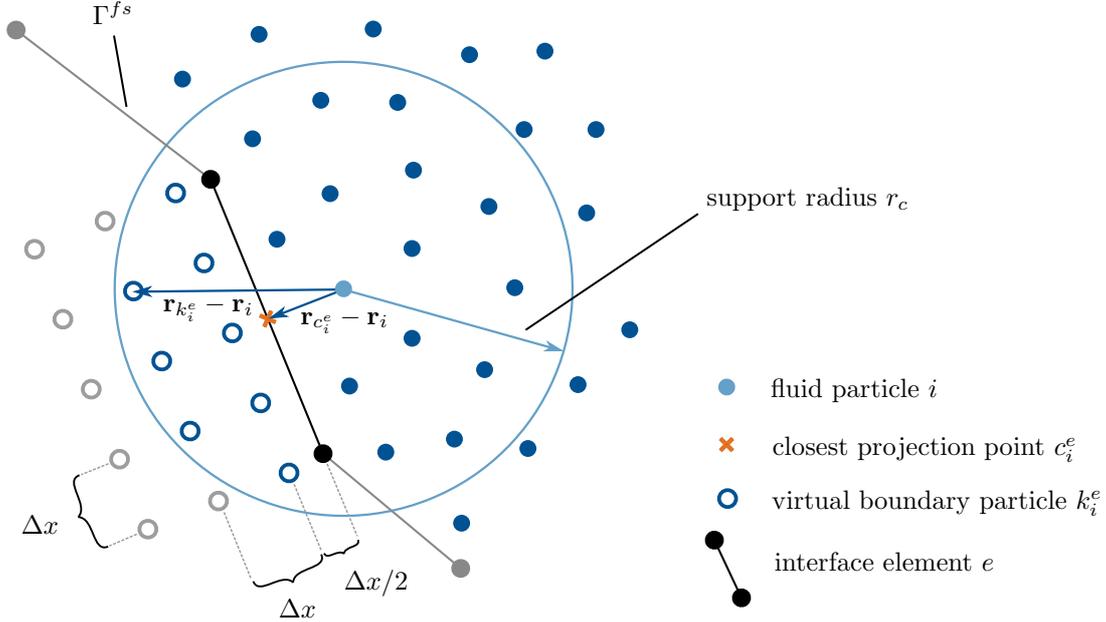}
\caption{Fluid particle~$i$ with closest projection point~$c^{e}_{i}$ to the interface~$\Gamma^{fs}$ on interface element~$e$ within support radius~$r_{c}$ of the smoothing kernel~$W$ and corresponding set of virtual boundary particles~$k^{e}_{i}$ associated with fluid particle~$i$ and ensuring full support of the smoothing kernel~$W$.}
\label{fig:inter_sketch}
\end{figure}

The interaction evaluation is performed between fluid particles and interface elements. Consider a fluid particle~$i$ with support radius~$r_{c}$ of the smoothing kernel~$W$ that is close to the fluid-structure interface~$\Gamma^{fs}$, cf. Figure~\ref{fig:inter_sketch}.

% begin figure
% end figure

In general, the closest projection point~$c^{e}_{i}$ of fluid particle~$i$ to the interface~$\Gamma^{fs}$ is located on interface element~$e$ and lies within the support radius, i.e., $\qty| \vectorbold{r}_{c^{e}_{i}} - \vectorbold{r}_{i} | < r_{c}$. The position~$\vectorbold{r}_{c^{e}_{i}}$ of point~$c^{e}_{i}$ can be described in iso-parametric coordinates~$\vectorbold*{\xi}_{c^{e}_{i}}$ on interface element~$e$. As a result, the shape functions~$N^{e}_{j}\qty\big(\vectorbold*{\xi}_{c^{e}_{i}})$ of all nodes~$j$ of interface element~$e$ evaluated at the closest projection point~$c^{e}_{i}$ can be utilized to interpolate kinematic quantities, e.g., positions, velocities, and accelerations, at the closest projection point using nodal quantities and to distribute kinetic quantities, e.g., interaction forces, from the closest projection point to adjacent nodes. The closest projection point~$c^{e}_{i}$ of a fluid particle~$i$ to a neighboring interface element~$e$ is detected solving the following minimization problem
\begin{equation}
\qty| \vectorbold{r}_{c^{e}_{i}} - \vectorbold{r}_{i} | = \min_{\vectorbold*{\xi}} \qty| \sum_{j} N^{e}_{j}\qty\big(\vectorbold*{\xi}) \, \vectorbold{r}^{fs}_{j} - \vectorbold{r}_{i} |
\end{equation}
with position~$\vectorbold{r}_{i}$ of fluid particle~$i$ and positions~$\vectorbold{r}^{fs}_{j}$ of nodes~$j$ of interface element~$e$. The solution of the minimization problem gives the iso-parametric coordinates~$\vectorbold*{\xi}_{c^{e}_{i}}$ of the closest projection point~$c^{e}_{i}$ on interface element~$e$. Hence, the position of the closest projection point~$c^{e}_{i}$ results in
\begin{equation} \label{eq:inter_pos_closestproj}
\vectorbold{r}_{c^{e}_{i}} = \sum_{j} N^{e}_{j}\qty\big(\vectorbold*{\xi}_{c^{e}_{i}}) \, \vectorbold{r}^{fs}_{j} \, .
\end{equation}
As stated above only closest projection points~$c^{e}_{i}$ located within the support radius of fluid particle~$i$ are considered in the interaction evaluation, meaning in addition $\qty| \vectorbold{r}_{c^{e}_{i}} - \vectorbold{r}_{i} | < r_{c}$ must be fulfilled. By definition, when evolving the position of a fluid particle~$i$ over time, also the position of the closest projection point~$c^{e}_{i}$ is changing, i.e., is sliding on the interface~$\Gamma^{fs}$.

\begin{figure}[htbp]
\centering
% begin subfigure
\subfigure [case 1: convex angle]
{
\newcommand*{\scalefig}{0.75}
\includegraphics[scale=\scalefig]{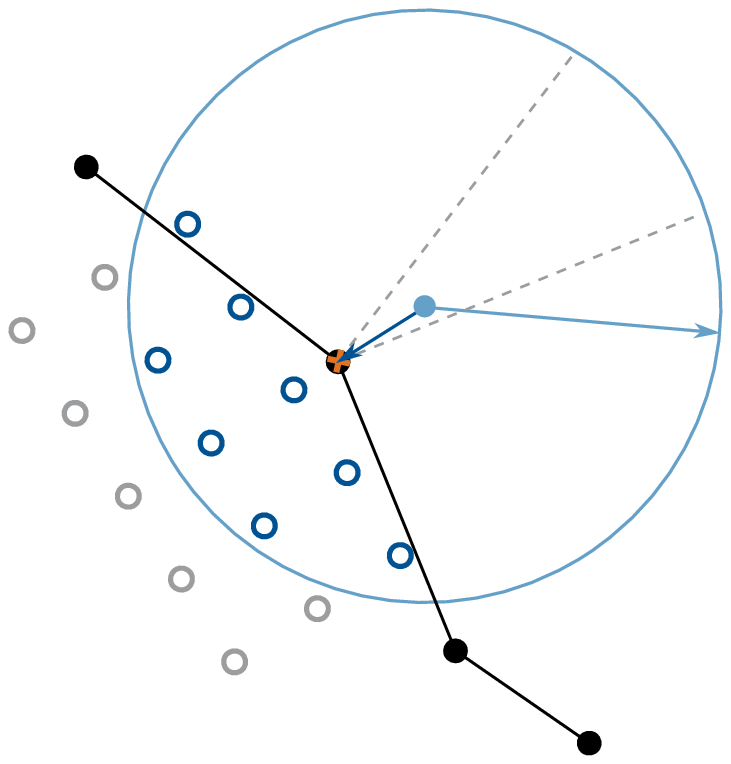}
}
% end subfigure
\hspace{0.1\textwidth}
% begin subfigure
\subfigure [case 2: concave angle]
{
\newcommand*{\scalefig}{0.75}
\includegraphics[scale=\scalefig]{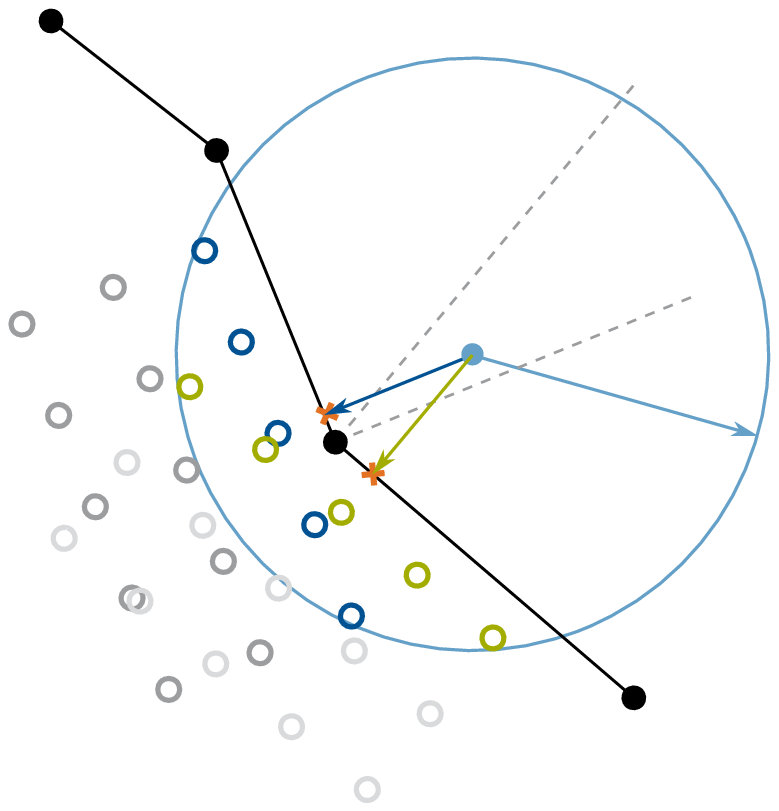}
}
% end subfigure
\caption{Special cases of a convex and a concave angle between neighboring interface elements for the treatment of closest projection points of a fluid particle, cf. Remark~\ref{rmk:inter_cases} (legend similar to Figure~\ref{fig:inter_sketch}).}
\label{fig:inter_cases}
\end{figure}

% begin remark
\begin{rmk} \label{rmk:inter_cases}
In the general case, the closest projection point of a particle is located on the surface of an interface element, as illustrated for instance in Figure~\ref{fig:inter_sketch}. In addition, the two special cases of a convex and a concave angle between two neighboring interface elements are worth being discussed here. In the case a particle is located within the perpendicular straight lines of neighboring interface elements at a convex angle, cf. \textit{case 1} in Figure~\ref{fig:inter_cases}, a single closest projection point is considered that is located on the node respectively the edge being shared by those interface elements. For a particle located at a concave angle, cf. \textit{case 2} in Figure~\ref{fig:inter_cases}, multiple closest projection points on the surface of each of the interface elements are considered.
\end{rmk}
% end remark

% begin figure
% end figure

% begin remark
\begin{rmk}
Note that very similar to typical contact problems in finite element analysis an extension to a $C^{1}$-continuous representation of the structural geometry, e.g., by employing Hermite polynomials~\cite{Meier2017a} or B-Splines~\cite{Hughes2005,Cottrell2009} as shape functions, could be beneficial within the proposed sliding boundary particle approach in terms of a smoother interaction force evolution and help to abstain from the aforementioned case distinctions, cf. Remark~\ref{rmk:inter_cases}.
\end{rmk}
% end remark

%%
\subsubsection{Virtual boundary particles} \label{subsec:nummeth_inter_virtbdrypart}

The support of the smoothing kernel of a fluid particle~$i$ close to the fluid-structure interface~$\Gamma^{fs}$ is truncated, i.e., fluid particle~$i$ experiences reduced contributions from neighboring fluid particles, cf. Figure~\ref{fig:inter_sketch}. To overcome this issue, full support of the smoothing kernel of fluid particle~$i$ is retained by considering a set of virtual boundary particles~$k^{e}_{i}$ that contribute to the interaction evaluation of fluid particle~$i$ and are regularly and equidistantly arranged behind the closest projection point~$c^{e}_{i}$ as illustrated in Figure~\ref{fig:inter_sketch}. This is achieved by a certain number of layers of virtual boundary particles with spacing~$\Delta{}x$ among them. Accordingly, together with the closest projection point~$c^{e}_{i}$, the set of virtual boundary particles~$k^{e}_{i}$ are sliding along the fluid-structure interface~$\Gamma^{fs}$ following the movement of a fluid particle~$i$, giving rise to the name of the proposed method: sliding boundary particle approach.

% begin remark
\begin{rmk}
Within this work, as stated in Sections~\ref{subsec:nummeth_sph_kernel} and~\ref{subsec:nummeth_sph_spacing}, a quintic spline smoothing kernel with support radius~$r_{c}=3h$ is applied with initial particle spacing~$\Delta{}x$ equal to the smoothing length~$h$. As a consequence, three layers of virtual boundary particles are positioned behind the closest projection point~$c^{e}_{i}$, thus, maintaining full support of the smoothing kernel~$W$ of fluid particle~$i$, cf. Figure~\ref{fig:inter_sketch}.
\end{rmk}
% end remark

All layers of virtual boundary particles are positioned perpendicular to the connection vector~$\vectorbold{r}_{c^{e}_{i}} - \vectorbold{r}_{i}$ of fluid particle~$i$ and its closest projection point~$c^{e}_{i}$, where the first layer is at a distance of~$\flatfrac{\Delta{}x}{2}$ behind the closest projection point~$c^{e}_{i}$ on interface element~$e$. An orthonormal basis $\qty( \vectorbold{e}_{r}, \vectorbold{e}_{s}, \vectorbold{e}_{t} )$ with first base vector $\vectorbold{e}_{r} = \flatfrac{\qty( \vectorbold{r}_{c^{e}_{i}} - \vectorbold{r}_{i} )}{\qty| \vectorbold{r}_{c^{e}_{i}} - \vectorbold{r}_{i} |}$ is constructed~\cite{Hughes1999}. Consequently, the position of all virtual boundary particles~$k^{e}_{i}$ can be given in terms of the particle spacing~$\Delta{}x$ and the constructed orthonormal basis $\qty( \vectorbold{e}_{r}, \vectorbold{e}_{s}, \vectorbold{e}_{t} )$ as
\begin{equation}
\vectorbold{r}_{k^{e}_{i}} = \vectorbold{r}_{c^{e}_{i}} + \qty( m_{r} + \flatfrac{1}{2} ) \, \Delta{}x \, \vectorbold{e}_{r} + m_{s} \, \Delta{}x \, \vectorbold{e}_{s} + m_{t} \, \Delta{}x \, \vectorbold{e}_{r}
\end{equation}
with integers $m_{r} \in \qty{ 0, 1, \dots, \qty(q-1) }$ and $m_{s}, m_{t} \in \qty{ -\qty(q-1), \dots, \qty(q-1) }$ where $q = \mathrm{floor}\qty(\flatfrac{r_{c}}{\Delta{}x})$ defines the number of particles necessary to maintain full support of the smoothing kernel. Finally, the vector from fluid particle $i$ to virtual boundary particle~$k^{e}_{i}$ is $\vectorbold{r}_{k^{e}_{i}} - \vectorbold{r}_{i}$, cf. Figure~\ref{fig:inter_sketch}.

% begin remark
\begin{rmk}
The $\mathrm{floor}$ operator used herein is defined by $\mathrm{floor}\qty(x) := \max\qty{ k \in \mathbb{Z} \mid k \leq x}$ and returns the largest integer that is less than or equal to its argument~$x$.
\end{rmk}
% end remark

%%
\subsubsection{Interaction forces on fluid particles} \label{subsec:nummeth_inter_evaluation}

A fluid particle~$i$ close to the fluid-structure interface~$\Gamma^{fs}$, i.e., for which the closest projection point~$c^{e}_{i}$ on interface element~$e$ is within the support radius~$\qty| \vectorbold{r}_{c^{e}_{i}} - \vectorbold{r}_{i} | < r_{c}$ of fluid particle~$i$, additionally experiences contributions to the density summation~\eqref{eq:sph_densum} and the momentum evaluation~\eqref{eq:sph_momentum} from all virtual boundary particles~$k^{e}_{i}$ for which $\qty| \vectorbold{r}_{k^{e}_{i}} - \vectorbold{r}_{i} | < r_{c}$ holds, cf. Figure~\ref{fig:inter_sketch}.

As described in Section~\ref{subsec:nummeth_sph_summation} the density field is computed via summation of the respective smoothing kernel contributions of neighboring fluid particles~$j$, refer to equation~\eqref{eq:sph_densum}. Hence, considering the additional contributions of virtual boundary particles~$k^{e}_{i}$, the density summation for a fluid particle~$i$ reads
\begin{equation}
\rho_{i} = m_{i} \sum_{j} W_{ij} + m_{i} \sum_{e} \sum_{k^{e}_{i}} W_{ik^{e}_{i}}
\end{equation}
ensuring full support of the smoothing kernel.

Inspired by the treatment of boundary particles for rigid walls~\cite{Adami2012} the properties of virtual boundary particles~$k^{e}_{i}$, i.e., density, pressure, and velocity, are extrapolated based on the corresponding quantities from neighboring fluid particles~$j$ of closest projection point~$c^{e}_{i}$ on interface element~$e$. The goal is to achieve an undisturbed pressure field of fluid particles close to the interface. Satisfying the kinematic coupling condition on the fluid-structure interface~$\Gamma^{fs}$, cf. equation \eqref{eq:coup_geom_kinem}, also called no-slip boundary condition, viscous forces are considered in the momentum equation. It shall be noted, that some boundary particle formulations in SPH are based on the assumption of zero normal pressure gradients close to the interface. However, in~\cite{Adami2012} it is shown, that including the pressure gradient obtained from a local force balance is beneficial to accurately model the pressure field of fluid particles close to the boundary. Therefore, a similar strategy is pursued in the following.

\begin{figure}[htbp]
\centering
\newcommand*{\scaletext}{1.0}
\newcommand*{\scalefig}{1.0}
\psfrag{gfs}{\scalebox{\scaletext}{$\Gamma^{fs}$}}
\psfrag{r_c}{\scalebox{\scaletext}{support radius~$r_c$}}
\psfrag{ap}{\scalebox{\scaletext}{averaged centroid position $\expval{\vectorbold{r}}_{f}$}}
\psfrag{fp}{\scalebox{\scaletext}{fluid particle~$i$}}
\psfrag{cp}{\scalebox{\scaletext}{closest projection point~$c^{e}_{i}$}}
\psfrag{vp}{\scalebox{\scaletext}{virtual boundary particle~$k^{e}_{i}$}}
\psfrag{ife}{\scalebox{\scaletext}{interface element~$e$}}
\psfrag{r_cf}{\scalebox{\scaletext}{$\vectorbold{r}_{c^{e}_{i}} - \expval{\vectorbold{r}}_{f}$}}
\psfrag{r_kf}{\scalebox{\scaletext}{$\vectorbold{r}_{k^{e}_{i}} - \expval{\vectorbold{r}}_{f}$}}
\includegraphics[scale=\scalefig]{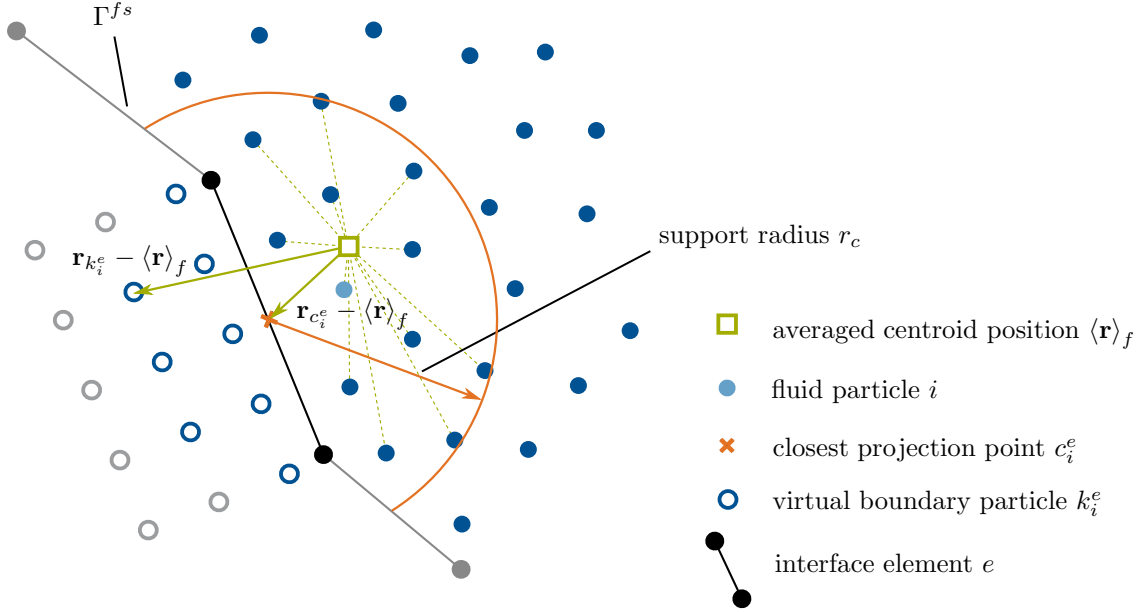}
\caption{Averaged centroid position~$f$ of the domain covered by neighboring fluid particles~$j$ of closest projection point~$c^{e}_{i}$.}
\label{fig:inter_extrapolation}
\end{figure}

In a first step, the pressure~$p_{k^{e}_{i}}$ of virtual boundary particles~$k^{e}_{i}$ is approximated based on a first order Taylor series expansion with center of expansion at
\begin{equation} \label{eq:inter_averagedcentroidpos}
\expval{\vectorbold{r}}_{f} = \frac{\sum_{j} \vectorbold{r}_{j} W_{c^{e}_{i}j}}{\sum_{j} W_{c^{e}_{i}j}} \, .
\end{equation}
The position~$\expval{\vectorbold{r}}_{f}$ can be interpreted as smoothed or averaged centroid position of the domain covered by the neighboring fluid particles $j$ as illustrated in Figure~\ref{fig:inter_extrapolation}. Hence, the pressure of virtual boundary particles~$k^{e}_{i}$ is determined following
\begin{equation} \label{eq:inter_virt_bdry_part_pressure}
p_{k^{e}_{i}} = \expval{p}_{f} + \qty( \vectorbold{r}_{k^{e}_{i}} - \expval{\vectorbold{r}}_{f} ) \vdot \expval{\grad{p}}_{f}
\end{equation}
with smoothed pressure $\expval{p}_{f} = \flatfrac{\sum_{j} p_{j} W_{c^{e}_{i}j}}{\sum_{j} W_{c^{e}_{i}j}}$ and smoothed pressure gradient
\begin{equation} \label{eq:inter_smoothedpressuregrad}
\expval{\grad{p}}_{f} = \frac{\sum_{j} \rho_{j} W_{c^{e}_{i}j}}{\sum_{j} W_{c^{e}_{i}j}} \qty(\vectorbold{b}_{i} - \vectorbold{a}_{c^{e}_{i}}) \, .
\end{equation}
The latter is approximated based on a local force balance neglecting viscous forces as proposed in~\cite{Adami2012}, cf. equation \eqref{eq:fluid_momentum}, with acceleration~$\vectorbold{a}_{c^{e}_{i}}$ of the closest projection point~$c^{e}_{i}$, cf. Remark~\ref{rmk:inter_velandacc_closestproj}. Applying the equation of state~\eqref{eq:sph_eos} of the respective interacting fluid particle~$i$ together with pressure~$p_{k^{e}_{i}}$, the density~$\rho_{k^{e}_{i}}$ of virtual boundary particles~$k^{e}_{i}$ follows as
\begin{equation}
\rho_{k^{e}_{i}} = \frac{p_{k^{e}_{i}}}{c^{2}} + \rho_{0} \, .
\end{equation}

% begin remark
\begin{rmk}
Note that the approximation of the smoothed pressure gradient $\expval{\grad{p}}_{f}$, cf. equation~\eqref{eq:inter_smoothedpressuregrad}, could be improved considering viscous forces in the local force balance, e.g.,~\cite{Hashemi2012}, however, at the cost of additional computational and algorithmic effort.
\end{rmk}
% end remark

% begin remark
\begin{rmk} \label{rmk:inter_velandacc_closestproj}
Similar to the position~$\vectorbold{r}_{c^{e}_{i}}$ of the closest projection point~$c^{e}_{i}$, cf. equation \eqref{eq:inter_pos_closestproj}, the velocity and acceleration are obtained following $\vectorbold{u}_{c^{e}_{i}} = \sum_{j} N^{e}_{j}\qty\big(\vectorbold*{\xi}_{c^{e}_{i}}) \, \vectorbold{u}^{fs}_{j}$ and $\vectorbold{a}_{c^{e}_{i}} = \sum_{j} N^{e}_{j}\qty\big(\vectorbold*{\xi}_{c^{e}_{i}}) \, \vectorbold{a}^{fs}_{j}$, where~$\vectorbold{u}^{fs}_{j}$ and~$\vectorbold{a}^{fs}_{j}$ are the velocities and accelerations of nodes~$j$ of interface element~$e$.
\end{rmk}
% end remark

% begin figure
% end figure

In a next step, the velocity~$\vectorbold{u}_{k^{e}_{i}}$ of virtual boundary particles~$k^{e}_{i}$ is approximated considering the kinematic coupling condition on the fluid-structure interface~$\Gamma^{fs}$, cf. equation \eqref{eq:coup_geom_kinem}, prescribing the velocity~$\vectorbold{u}_{c^{e}_{i}}$ of closest projection point~$c^{e}_{i}$, cf. Remark~\ref{rmk:inter_velandacc_closestproj}. Applying a first order Taylor series expansion with center of expansion at $\expval{\vectorbold{r}}_{f}$ according to \eqref{eq:inter_averagedcentroidpos} gives the relation
\begin{equation} \label{eq:inter_taylorexpansion_uc}
\vectorbold{u}_{c^{e}_{i}} = \expval{\vectorbold{u}}_{f} + \qty( \vectorbold{r}_{c^{e}_{i}} - \expval{\vectorbold{r}}_{f} ) \vdot \vectorbold{e}_{r} \expval{\grad_{\vectorbold{e}_{r}}{\vectorbold{u}}}_{f}
\end{equation}
with smoothed velocity $\expval{\vectorbold{u}}_{f} = \flatfrac{\sum_{j} \vectorbold{u}_{j} W_{c^{e}_{i}j}}{\sum_{j} W_{c^{e}_{i}j}}$ and unit vector $\vectorbold{e}_{r}$ pointing from particle~$i$ to closest projection point~$c^{e}_{i}$ thus representing the wall normal vector as defined in Section~\ref{subsec:nummeth_inter_virtbdrypart}. The quantity $\expval{\grad_{\vectorbold{e}_{r}}{\vectorbold{u}}}_{f}$ denotes the smoothed directional derivative of the velocity in direction of~$\vectorbold{e}_{r}$ and follows from equation~\eqref{eq:inter_taylorexpansion_uc} as
\begin{equation}
\expval{\grad_{\vectorbold{e}_{r}}{\vectorbold{u}}}_{f} = \frac{\vectorbold{u}_{c^{e}_{i}} - \expval{\vectorbold{u}}_{f}}{\qty( \vectorbold{r}_{c^{e}_{i}} - \expval{\vectorbold{r}}_{f} ) \vdot \vectorbold{e}_{r}}
\end{equation}
exploiting the velocity~$\vectorbold{u}_{c^{e}_{i}}$ of the closest projection point~$c^{e}_{i}$. Finally, the velocity~$\vectorbold{u}_{k^{e}_{i}}$ of virtual boundary particles~$k^{e}_{i}$ is approximated again applying a Taylor series expansion with center of expansion at $\expval{\vectorbold{r}}_{f}$ resulting in
\begin{equation} \label{eq:inter_virt_bdry_part_velocity}
\vectorbold{u}_{k^{e}_{i}} = \expval{\vectorbold{u}}_{f} + \qty( \vectorbold{r}_{k^{e}_{i}} - \expval{\vectorbold{r}}_{f} ) \vdot \vectorbold{e}_{r} \expval{\grad_{\vectorbold{e}_{r}}{\vectorbold{u}}}_{f} \, .
\end{equation}
In addition to the acceleration contributions~$\vectorbold{a}_{ij}$ of neighboring fluid particles~$j$, the momentum equation~\eqref{eq:sph_momentum} for a fluid particle~$i$ is extended by the acceleration contributions~$\vectorbold{a}_{ik^{e}_{i}}$ of virtual boundary particles~$k^{e}_{i}$ related to the closest projection points~$c^{e}_{i}$ on interface elements~$e$
\begin{equation}
\vectorbold{a}_{i} = \sum_{j} \vectorbold{a}_{ij} + \sum_{e} \sum_{k^{e}_{i}} \vectorbold{a}_{ik^{e}_{i}} + \vectorbold{b}_{i}
\end{equation}
with
\begin{equation}
\vectorbold{a}_{ik^{e}_{i}} = \frac{1}{m_{i}} \qty(V_{i}^{2}+V_{k^{e}_{i}}^{2}) \qty[ - \tilde{p}_{ik^{e}_{i}} \pdv{W}{r_{ik^{e}_{i}}} \vectorbold{e}_{ik^{e}_{i}} + \eta_{i} \frac{\vectorbold{u}_{ik^{e}_{i}}}{r_{ik^{e}_{i}}} \pdv{W}{r_{ik^{e}_{i}}} ]
\end{equation}
and density-weighted inter-particle averaged pressure~$\tilde{p}_{ik^{e}_{i}}$ as defined in~\eqref{eq:sph_mom_wght_press}.

% begin remark
\begin{rmk} \label{rmk:inter_advantage_computational}
The extrapolation of pressure and velocity for virtual boundary particles $k^{e}_{i}$ by the Taylor series expansions \eqref{eq:inter_virt_bdry_part_pressure} and \eqref{eq:inter_virt_bdry_part_velocity} requires the quantities $\expval{\cdot}_{f}$ to be evaluated only once for each closest projection point $c^{e}_{i}$, which is the main advantage of this procedure regarding computational costs.
\end{rmk}
% end remark

% begin remark
\begin{rmk}
Note that the contributions of virtual boundary particles~$k^{e}_{i}$ resulting from a background pressure~$p_{b}$ as part of the transport velocity formulation~\cite{Adami2013} are also considered for fluid particles~$i$, however, similar to Section~\ref{subsec:nummeth_sph_momentum} for ease of notation not pointed out here.
\end{rmk}
% end remark

%%
\subsubsection{Nodal interface forces on interface elements} \label{subsec:nummeth_inter_nodalforce}

The coupling of the fluid and the structural field, cf. Figure~\ref{fig:domain_discretized}, following a Dirichlet-Neumann partitioned approach (as discussed in more detail in Section~\ref{subsec:nummeth_coupappr}) requires the evaluation of interface forces~$\vectorbold{f}^{fs}$. To enforce conservation of linear momentum at the fluid-structure interface~$\Gamma^{fs}$, in accordance with~\eqref{eq:sph_conservlinmom}, the interface forces~$\vectorbold{f}^{fs}$ can be computed directly from the resulting acceleration contributions of fluid particles interacting with virtual boundary particles, as described in the previous section. Consequently, the resulting force~$\vectorbold{f}^{e}_{c^{e}_{i}}$ acting on an interface element~$e$ at the closest projection point~$c^{e}_{i}$ due to interaction of virtual boundary particles~$k^{e}_{i}$ with fluid particle~$i$ reads
\begin{equation} \label{eq:inter_resforce}
\vectorbold{f}^{e}_{c^{e}_{i}} = - m_{i} \sum_{k^{e}_{i}} \vectorbold{a}_{ik^{e}_{i}}
\end{equation}
with mass~$m_{i}$ of fluid particle~$i$ and acceleration contribution~$\vectorbold{a}_{ik^{e}_{i}}$ of virtual boundary particle~$k^{e}_{i}$ on fluid particle~$i$. Note that the above given formulation of the resulting force~$\vectorbold{f}^{e}_{c^{e}_{i}}$ guarantees conservation of linear momentum between fluid particle~$i$ and interface element~$e$. The resulting force~$\vectorbold{f}^{e}_{c^{e}_{i}}$ (being a point force acting on interface element~$e$ at closest projection point~$c^{e}_{i}$) is distributed to the nodes~$j$ of interface element~$e$ using its shape functions~$N^{e}_{j}\qty\big(\vectorbold*{\xi}_{c^{e}_{i}})$ evaluated at the closest projection point~$c^{e}_{i}$ given in iso-parametric coordinates~$\vectorbold*{\xi}_{c^{e}_{i}}$. Finally, the interface force~$\vectorbold{f}^{fs}_{j}$ of a node~$j$ results from summation over all force contributions~$\vectorbold{f}^{e}_{c^{e}_{i}}$ of fluid particles~$i$ acting on various interface elements~$e$ connected to node~$j$
\begin{equation}
\vectorbold{f}^{fs}_{j} = \sum_{e} \sum_{i} N^{e}_{j}\qty\big(\vectorbold*{\xi}_{c^{e}_{i}}) \vectorbold{f}^{e}_{c^{e}_{i}} \, .
\end{equation}

\subsection{Partitioned coupling approach} \label{subsec:nummeth_coupappr}

The fluid and the structural field are coupled following a Dirichlet-Neumann partitioned approach, where the fluid field is the Dirichlet partition with prescribed interface displacements~$\vectorbold{d}^{fs}$ and the structural field is the Neumann partition subject to interface forces~$\vectorbold{f}^{fs}$, as illustrated in Figure~\ref{fig:domain_discretized} (right).

Introducing the field operators~$\mathcal{F}$ and~$\mathcal{S}$ for the fluid and the structural problem~\cite{Kuttler2008} both mapping the interface displacements~$\vectorbold{d}^{fs}$ to interface forces
\begin{equation}
\vectorbold{f}^{fs}_{\mathcal{F}} = \mathcal{F}\qty(\vectorbold{d}^{fs}) \qand \vectorbold{f}^{fs}_{\mathcal{S}} = \mathcal{S}\qty(\vectorbold{d}^{fs}) \, ,
\end{equation}
equilibrium at the interface~$\Gamma^{fs}$ is satisfied in case the condition
\begin{equation}
\mathcal{F}\qty(\vectorbold{d}^{fs}) = \mathcal{S}\qty(\vectorbold{d}^{fs})
\end{equation}
holds. The inverse fluid and structural field operators mapping interface forces~$\vectorbold{f}^{fs}$ to interface displacements are consequently defined as
\begin{equation}
\vectorbold{d}^{fs}_{\mathcal{F}} = \mathcal{F}^{-1}\qty(\vectorbold{f}^{fs}) \qand \vectorbold{d}^{fs}_{\mathcal{S}} = \mathcal{S}^{-1}\qty(\vectorbold{f}^{fs}) \, .
\end{equation}

In~\cite{Forster2007} it is shown that weakly coupled schemes exhibit instabilities in FSI problems with incompressible flows due to the artificial added mass effect. To overcome those instabilities, a fixed-point coupling algorithm is employed to iteratively reach dynamic equilibrium of the fluid and the structural field at the interface with respect to a predefined convergence criterion, i.e., fluid and structural field are strongly coupled. Following a synchronous time stepping scheme the same time step size~$\Delta{}t$ is set for both fluid and structural solver and is based on the in general more severe restrictions of the SPH time integration scheme, cf. equation~\eqref{eq:sph_timestepcond}.

% begin remark
\begin{rmk} \label{rmk:coup_asynchronous_time_stepping}
Note that the applied generalized alpha time integration scheme for the structural field being an implicit method in general allows for a larger time step size~$\Delta{}t$ than possible for the fluid field solved using SPH. Thus, future research may focus on asynchronous time stepping and sub-stepping schemes in order to reduce computational costs.
\end{rmk}
% end remark

The coupling algorithm applied herein is described in detail below as Algorithm~\ref{alg:part_fsi}. Convergence of the iterative coupling loop in Algorithm~\ref{alg:part_fsi} is achieved in case the following criterion based on the increment of interface displacements~$\Delta\vectorbold{d}^{fs}_{n+1,i+1}$ is fulfilled
\begin{equation} \label{eq:coup_convergence}
\frac{ \qty\big|\Delta\vectorbold{d}^{fs}_{n+1,i+1}|}{\Delta{}t \, \sqrt{n^{fs}_{dof}}} < \epsilon
\end{equation}
with time step size~$\Delta{}t$, number of interface degrees of freedom~$n^{fs}_{dof}$, and predefined tolerance for convergence~$\epsilon$.

% begin remark
\begin{rmk} \label{rmk:coup_aitken}
In general, applying dynamic relaxation of the interface displacements~$\vectorbold{d}^{fs}$ in each iteration of the coupling algorithm~\cite{Kuttler2008} can have a stabilizing effect and accelerate the convergence of the partitioned coupling. However, it shall be noted, that due to the restrictions of the time step size~$\Delta{}t$ resulting from the SPH time integration scheme, an accelerating effect is not required with the proposed formulation, cf. examples~\ref{subsec:numex_flexiblebeam} and~\ref{subsec:numex_academic_balloon}.
\end{rmk}
% end remark

% begin algorithm
\begin{algorithm}[htbp]
\caption{Time loop of a Dirichlet-Neumann partitioned fixed-point fluid-structure interaction algorithm} \label{alg:part_fsi}
\begin{algorithmic}[0]\onehalfspacing
\While{$t < T$}
%	\Statex
	\State $t \gets t + \Delta{T}$
	\Comment increment time
	\State $i \gets 1$
	\Comment reset iteration counter
%	\Statex
	\State $\vectorbold{d}^{fs}_{n+1,i}$
	\Comment predict interface displacements
%	\Statex
	\While{$true$}
%		\Statex
		\State $\vectorbold{f}^{fs}_{n+1,i+1} = \mathcal{F}\qty(\vectorbold{d}^{fs}_{n+1,i})$
		\Comment solve fluid field
%		\Statex
		\State $\vectorbold{d}^{fs}_{n+1,i+1} = \mathcal{S}^{-1}\qty(\vectorbold{f}^{fs}_{n+1,i+1})$
		\Comment solve structural field
%		\Statex
		\State $\Delta\vectorbold{d}^{fs}_{n+1,i+1} = \vectorbold{d}^{fs}_{n+1,i+1} - \vectorbold{d}^{fs}_{n+1,i}$
		\Comment compute increment of interface displacements
%		\Statex
		\If {$\flatfrac{ \qty\big|\Delta\vectorbold{d}^{fs}_{n+1,i+1}|}{\Delta{}t \, \sqrt{n^{fs}_{dof}}} < \epsilon$}
		\Comment check convergence criterion, cf. equation~\eqref{eq:coup_convergence}
			\State \textbf{break}
		\EndIf
%		\Statex
		\State $i \gets i + 1$
		\Comment increment iteration counter
%		\Statex
	\EndWhile
%	\Statex
	\State $n \gets n + 1$
	\Comment increment step counter
%	\Statex
\EndWhile
\end{algorithmic}
\end{algorithm}
% end algorithm

%%
\section{Numerical examples} \label{sec:numex}

The purpose of this section is to validate the novel sliding boundary particle approach and the proposed numerical formulation for solving fluid-structure interaction problems examining several numerical examples in two and three dimensions. The obtained results are assessed on the basis of analytical solutions and reference solutions given in the literature.

\subsection{Validation of the sliding boundary particle approach}

At first, the capabilities of the proposed method considering fluid flow in the presence of rigid and undeformable structures with a focus on the validation of the novel sliding boundary particle approach as presented in Section~\ref{subsec:nummeth_inter} are shown. The obtained results are compared to analytical solutions and reference solutions given in the literature both in a quantitative and qualitative manner. Additionally, as rigid and undeformable structures are considered, these examples can also be examined utilizing an implementation of the rigid wall boundary condition proposed in~\cite{Adami2012}. As a result, this allows for validation of the proposed sliding boundary particle approach against the established rigid wall boundary condition~\cite{Adami2012} within the context of rigid and undeformable structures. Finally, an example demonstrates the advantages of the proposed sliding boundary particle approach in the regime of large structural deformations. In all examples discussed in this section, the structural field is not solved, though, the fluid-structure interface is explicitly described either via an interface mesh or analytically by parameterization.

\subsubsection{Hydrostatic pressure in a fluid between two parallel plates} \label{subsec:numex_hydrostatic_pressure}

\begin{figure}[htbp]
\centering
% begin subfigure
\subfigure [Particle distribution colored with fluid pressure ranging from $-0.01$ (blue) to $0.01$ (red) with illustration of coordinate axis and body force.]
{
\newcommand*{\scaletext}{1.0}
\psfrag{cq}{\scalebox{\scaletext}{$q$}}
\psfrag{bq}{\scalebox{\scaletext}{$b_{q}$}}
\includegraphics[width=0.4\textwidth]{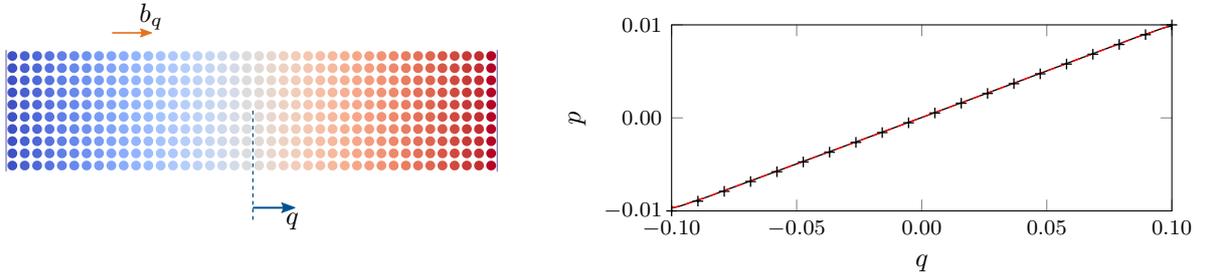}
\label{fig:example_hydrostatic_pressure_distribution}
}
% end subfigure
\hspace{0.02\textwidth}
% begin subfigure
\subfigure [Numerical result using the proposed sliding boundary particle approach (black solid line) and the rigid wall boundary condition~\citep{Adami2012} (red dashed line) compared to analytical solution (crosses).]
{
\hspace{0.08\textwidth}
\begin{tikzpicture}[trim axis left,trim axis right]
\begin{axis}
[
width=0.4\textwidth,
height=0.15\textwidth,
xmin=-0.1, xmax=0.1,
ymin=-0.01, ymax=0.01,
scaled ticks=false,
xticklabel style={/pgf/number format/precision=2, /pgf/number format/fixed, /pgf/number format/fixed zerofill},
ytick={-0.01, 0.00, 0.01},
yticklabel style={/pgf/number format/precision=2, /pgf/number format/fixed, /pgf/number format/fixed zerofill},
xlabel={$q$},
ylabel={$p$},
]
\addplot [color=black,only marks,mark=+,domain=-0.1:0.1,samples=20] {0.1*x};
\addplot [color=black,solid,line width=0.5pt] table [x expr={(\thisrow{"Points_X"})}, y expr={(\thisrow{"pressure_shepard_wall"})}, col sep=comma] {data/hydrostatic_pressure.csv};
\addplot [color=red,densely dashed,line width=0.5pt] table [x expr={(\thisrow{"Points_X"})}, y expr={(\thisrow{"pressure_shepard_bdry"})}, col sep=comma] {data/hydrostatic_pressure.csv};
\end{axis}
\end{tikzpicture}
\hspace{0.02\textwidth}
\label{fig:example_hydrostatic_pressure_comparison}
}
% end subfigure
\caption{Hydrostatic pressure in a fluid between two parallel plates: fluid pressure in static equilibrium state at time $t = 10.0$.}
\label{fig:example_hydrostatic_pressure}
\end{figure}

The gap between two spatially fixed and undeforming parallel plates being a distance of $L=0.2$ apart is filled with a Newtonian fluid of density~$\rho^{f} = 1.0$ and kinematic viscosity~$\nu^{f} = 1.0 \times 10^{-2}$. A coordinate axis~$\vectorbold{e}_{q}$ is introduced pointing in the direction perpendicular to the parallel plates with origin centered between the latter, cf. Figure~\ref{fig:example_hydrostatic_pressure_distribution}. Finally, a body force of magnitude $b_{q} = 0.1$ acting in direction~$\vectorbold{e}_{q}$ is applied on the fluid. For this simple example the analytical solution for the pressure profile in the static equilibrium state is given to $p\qty(q) = \rho^{f} b_{q} q$ showing linear behavior.

The fluid domain between the two parallel plates is discretized by 40 layers of fluid particles, i.e., with an initial particle spacing~$\Delta{}x = 5.0 \times 10^{-3}$. The smoothing length~$h$ of the smoothing kernel is set equal to the initial particle spacing~$\Delta{}x$ resulting in a support radius~$r_{c} = 1.5 \times 10^{-2}$. For the fluid phase, an artificial speed of sound~$c = 1.0$ is chosen, leading to a reference pressure~$p_{0} = 1.0$. The background pressure~$p_{b}$ is set equal to the reference pressure~$p_{0}$. The two parallel plates are modeled by a surface element each. The problem is solved with time step size~$\Delta{}t = 3.125 \times 10^{-4}$, cf. equation~\eqref{eq:sph_timestepcond}, until a static equilibrium state is reached.

% begin figure
% end figure

% begin figure
\begin{figure}[htbp]
\centering
\begin{tikzpicture}[trim axis left,trim axis right]
\begin{axis}
[
width=0.4\textwidth,
height=0.15\textwidth,
xmin=0.08, xmax=0.12,
ymin=0.008, ymax=0.012,
scaled ticks=false,
xticklabel style={/pgf/number format/precision=2, /pgf/number format/fixed, /pgf/number format/fixed zerofill},
ytick={0.008, 0.01, 0.012},
yticklabel style={/pgf/number format/precision=3, /pgf/number format/fixed, /pgf/number format/fixed zerofill},
xlabel={$q$},
ylabel={$p$},
]
\addplot [color=red,only marks,mark=o] table [x expr={(\thisrow{"q_coord"})}, y expr={(\thisrow{"boundary pressure"})}, col sep=comma] {data/hydrostatic_pressure_bdry_particle_bdry.csv};
\addplot [color=black,only marks,mark=o] table [x expr={(\thisrow{"q_coord"})}, y expr={(\thisrow{"boundary pressure"})}, col sep=comma] {data/hydrostatic_pressure_bdry_particle_wall.csv};
\addplot [color=blue,only marks,mark=o] table [x expr={(\thisrow{"q_coord"})}, y expr={(\thisrow{"boundary pressure"})}, col sep=comma] {data/hydrostatic_pressure_bdry_particle_wall_zeropressgrad.csv};
\addplot [color=red,only marks,mark=o] table [x expr={(\thisrow{"q_coord"})}, y expr={(\thisrow{"pressure"})}, col sep=comma] {data/hydrostatic_pressure_fluid_particle_bdry.csv};
\addplot [color=black,only marks,mark=o] table [x expr={(\thisrow{"q_coord"})}, y expr={(\thisrow{"pressure"})}, col sep=comma] {data/hydrostatic_pressure_fluid_particle_wall.csv};
\addplot [color=blue,only marks,mark=o] table [x expr={(\thisrow{"q_coord"})}, y expr={(\thisrow{"pressure"})}, col sep=comma] {data/hydrostatic_pressure_fluid_particle_wall_zeropressgrad.csv};
%
%\addplot [color=black,only marks,mark=+,domain=-0.1:0.1,samples=80] {0.1*x};
\addplot [color=black,solid,line width=0.5pt,domain=-0.1:0.1,samples=20] {0.1*x};
\addplot [color=gray,solid,line width=0.5pt] coordinates{(0.1,0.008) (0.1,0.012)};
\draw[draw=none, pattern=north east lines, pattern color=gray] (axis cs:0.1,0.008) rectangle (axis cs:0.12,0.012);
\end{axis}
\end{tikzpicture}
\caption{Hydrostatic pressure in a fluid between two parallel plates: detailed view of boundary region with pressure values of fluid particles and (virtual) boundary particles using the proposed sliding boundary particle approach \textit{with} (black circles) and \textit{without} (blue circles) considering the pressure gradient~\eqref{eq:inter_smoothedpressuregrad} in equation~\eqref{eq:inter_virt_bdry_part_pressure}, and the rigid wall boundary condition~\cite{Adami2012} also considering the pressure gradient (red circles) compared to analytical solution (black solid line).}
\label{fig:example_hydrostatic_pressure_particle}
\end{figure}
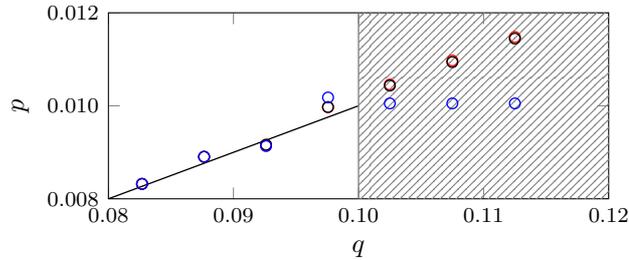
% end figure

Figure~\ref{fig:example_hydrostatic_pressure_comparison} shows the hydrostatic pressure in the fluid at time $t = 10.0$. The results compared to the analytical solution are post-processed applying SPH approximation~\eqref{eq:sph_postprocessing}. The results are in very good agreement with the analytical solution showing the capability of the proposed sliding boundary particle approach to capture linear pressure profiles near the boundary, cf. equation \eqref{eq:inter_virt_bdry_part_pressure}. In addition, the example is computed with an implementation of the rigid wall boundary condition~\cite{Adami2012} modeling the two parallel plates via fixed boundary particles. Comparing the result to those obtained with the proposed sliding boundary particle approach, cf. Figure~\ref{fig:example_hydrostatic_pressure_comparison}, delivers apart from roundoff errors equivalent results for this example. Finally, a detailed view of the boundary region at $q = 0.1$ is given in Figure~\ref{fig:example_hydrostatic_pressure_particle} showing the pressure values of fluid particles and (virtual) boundary particles obtained with the proposed sliding boundary particle approach and the rigid wall boundary condition~\cite{Adami2012}. In addition, a modified variant of the sliding boundary particle approach \textit{without} considering the pressure gradient~\eqref{eq:inter_smoothedpressuregrad} in equation~\eqref{eq:inter_virt_bdry_part_pressure} is examined. With this modified variant, the pressure value of the fluid particle closest to the boundary, cf. Figure~\ref{fig:example_hydrostatic_pressure_particle}, clearly deviates from the expected linear pressure profile. Thus, considering the improved accuracy and the fact that the computational costs required for the extrapolation of pressure (and velocity) for virtual boundary particles are negligible, cf. Remark~\ref{rmk:inter_advantage_computational}, in the following, the standard variant as proposed in Section~\ref{subsec:nummeth_inter_evaluation} is applied.

\subsubsection{Planar Taylor-Couette flow} \label{subsec:numex_taylorcouette}

\begin{figure}[htbp]
\centering
% begin subfigure
\subfigure [Geometry and boundary conditions of the problem.]
{
\newcommand*{\scaletext}{1.0}
\newcommand*{\scalefig}{0.5}
\psfrag{cr}{\scalebox{\scaletext}{$r$}}
\psfrag{ct}{\scalebox{\scaletext}{$\theta$}}
\psfrag{r1}{\scalebox{\scaletext}{$r_{1}$}}
\psfrag{r2}{\scalebox{\scaletext}{$r_{2}$}}
\psfrag{o1}{\scalebox{\scaletext}{$\omega_{1}$}}
\psfrag{o2}{\scalebox{\scaletext}{$\omega_{2}$}}
\psfrag{nosl}{\scalebox{\scaletext}{no-slip b.c.}}
\includegraphics[scale=\scalefig]{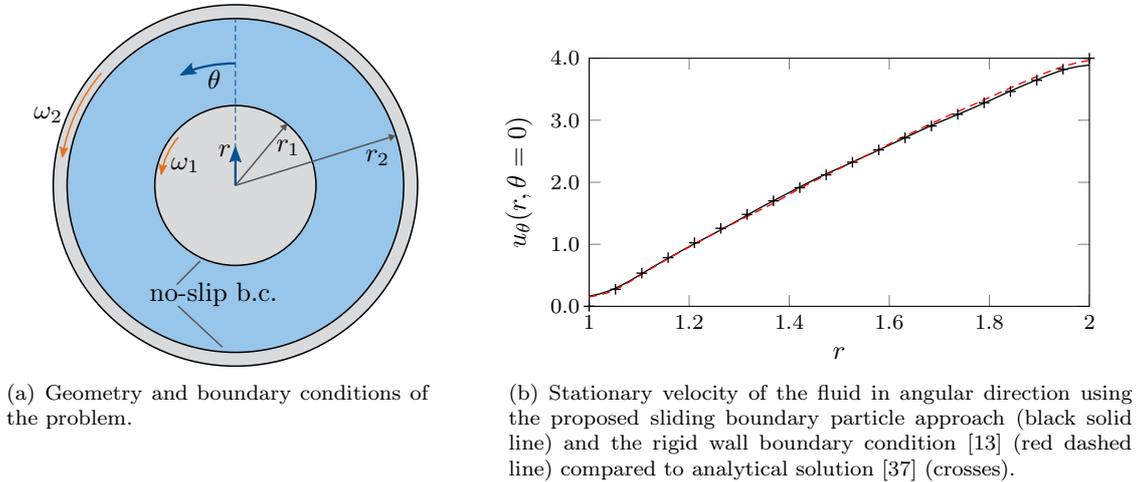}
\label{fig:example_taylorcouette_geometry}
}
% end subfigure
\hspace{0.05\textwidth}
% begin subfigure
\subfigure [Stationary velocity of the fluid in angular direction using the proposed sliding boundary particle approach (black solid line) and the rigid wall boundary condition~\cite{Adami2012} (red dashed line) compared to analytical solution~\cite{Limache2008} (crosses).]
{
\hspace{0.05\textwidth}
\begin{tikzpicture}[trim axis left,trim axis right]
\begin{axis}
[
width=0.4\textwidth,
height=0.2\textwidth,
xmin=1.0, xmax=2.0,
ymin=0.0, ymax=4.0,
scaled ticks=false,
yticklabel style={/pgf/number format/precision=1, /pgf/number format/fixed, /pgf/number format/fixed zerofill},
xlabel={$r$},
ylabel={$u_{\theta}\qty(r,\theta=0)$},
]
\addplot [color=black,only marks,mark=+,domain=1.0:2.0,samples=20] {-(8.0/3.0)*(1.0/x)+(8.0/3.0)*x};
\addplot [color=black,solid,line width=0.5pt] table [x expr={(\thisrow{"Points_Y"})}, y expr={(-1.0)*(\thisrow{"velocity_shepard_X"})}, col sep=comma] {data/taylorcouetteflow_vtheta_wall_circle.csv};
\addplot [color=red,densely dashed,line width=0.5pt] table [x expr={(\thisrow{"Points_Y"})}, y expr={(-1.0)*(\thisrow{"velocity_shepard_X"})}, col sep=comma] {data/taylorcouetteflow_vtheta_bdry.csv};
\end{axis}
\end{tikzpicture}
\hspace{0.02\textwidth}
\label{fig:example_taylorcouette_angular_vel}
}
% end subfigure
\caption{Planar Taylor-Couette flow: setup of the problem and numerical results.}
\label{fig:example_taylorcouette}
\end{figure}

In this example, a laminar, planar Taylor-Couette flow is considered. The gap between two coaxial cylinders with radii $r_{1} = 1.0$ and $r_{2} = 2.0$ is filled with a Newtonian fluid of density~$\rho^{f} = 1.0$ and kinematic viscosity~$\nu^{f} = 1.0$. The inner cylinder is fixed, i.e., its angular velocity is $\omega_{1} = 0.0$, while the outer cylinder rotates with angular velocity $\omega_{2} = 2.0$ around its axis of symmetry. No-slip boundary conditions are applied between the fluid and the surfaces of the cylinders. The geometry and boundary conditions of the problem are illustrated in Figure~\ref{fig:example_taylorcouette_geometry}. The Reynolds number of the problem is $Re = \flatfrac{ \omega_{2} r_{2} \qty(r_{2} - r_{1})}{\nu^{f}} = 4.0$ with maximum velocity $\omega_{2} r_{2}$ and gap $\qty(r_{2} - r_{1})$ between the coaxial cylinders.

% begin figure
% end figure

The fluid domain is discretized by fluid particles with initial particle spacing~$\Delta{}x = 5.0 \times 10^{-2}$. The smoothing length~$h$ is equal to the initial particle spacing~$\Delta{}x$ resulting in a support radius~$r_{c} = 1.5 \times 10^{-1}$. The artificial speed of sound is set to $c = 40.0$, hence the reference pressure is $p_{0} = 1600.0$. The background pressure~$p_{b}$ is set equal to the reference pressure~$p_{0}$. In this example, the structural surfaces are described analytically by parameterization of the cylindrical surfaces in order to show the capabilities and flexibility of the proposed sliding boundary particle approach. However, it shall be noted that the geometry naturally could have been discretized by a finite element mesh. The problem is solved with time step size~$\Delta{}t = 3.125 \times 10^{-4}$, cf. equation~\eqref{eq:sph_timestepcond}, until a nearly stationary state is reached at time $t=2.0$.

\begin{figure}[htbp]
\centering
% begin subfigure
\subfigure [Sliding boundary particle approach with parameterization of the cylindrical shape.]
{
\includegraphics[width=0.35\textwidth]{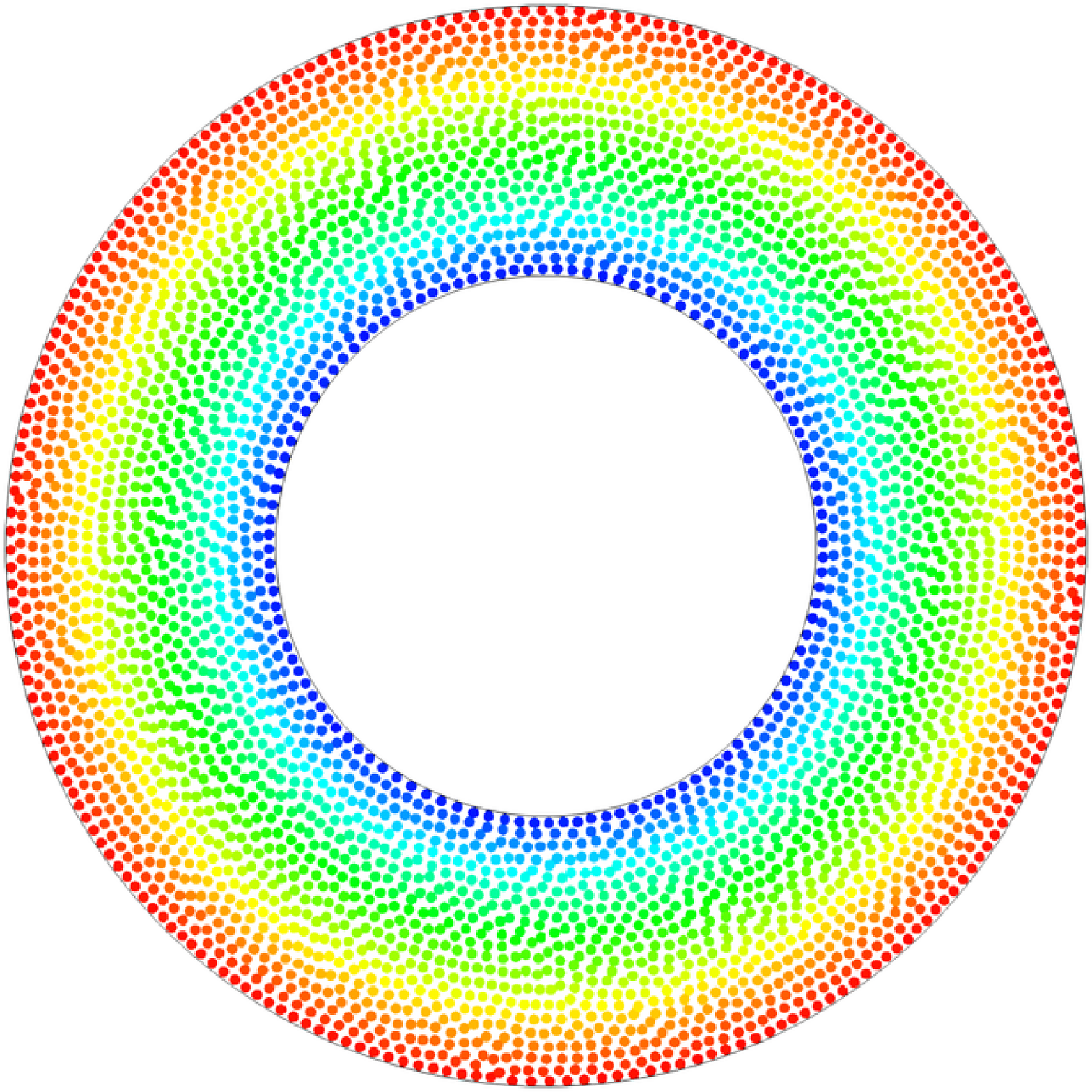}
}
% end subfigure
\hspace{0.1\textwidth}
% begin subfigure
\subfigure [Rigid wall boundary condition~\cite{Adami2012} with fixed boundary particles approximating the cylindrical shape.]
{
\includegraphics[width=0.35\textwidth]{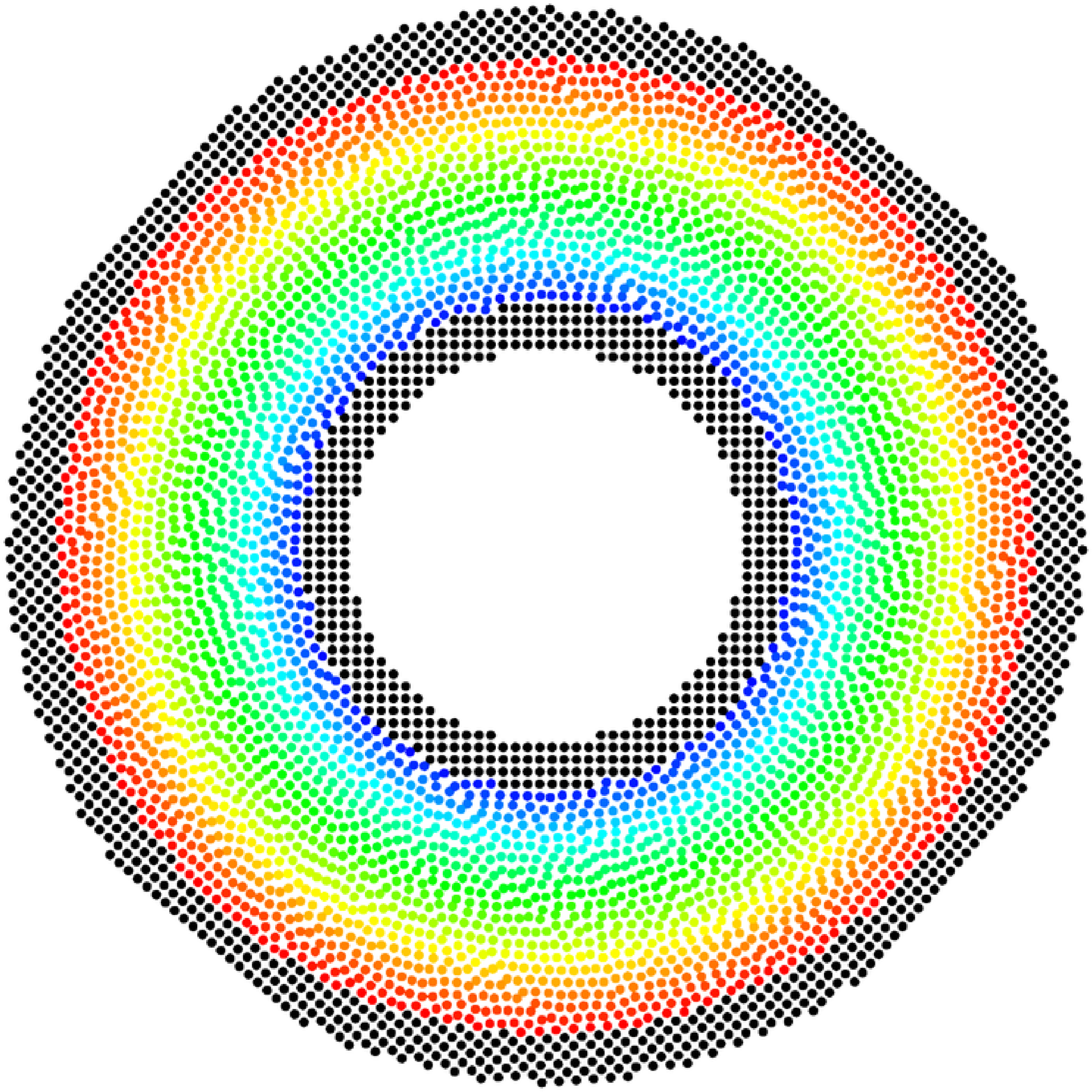}
}
% end subfigure
\caption{Planar Taylor-Couette flow: particle distribution at time $t = 2.0$ colored with magnitude of the fluid velocity ranging from $0.0$ (blue) to $4.0$ (red).}
\label{fig:example_taylorcouette_particle_distribution}
\end{figure}

In Figure~\ref{fig:example_taylorcouette_angular_vel} the stationary velocity of the fluid in the gap between the cylinders, post-processed applying SPH approximation~\eqref{eq:sph_postprocessing}, is plotted over the radius $r$ in angular direction $\theta$ at time $t = 2.0$. The result obtained with the proposed sliding boundary particle approach is compared to the result obtained with an implementation of the rigid wall boundary condition~\cite{Adami2012} and to the analytical solution of the problem~\cite{Limache2008}. Both methods show a high degree of conformity with the analytical solution. The deviation of the velocity in Figure~\ref{fig:example_taylorcouette_angular_vel} close to the cylindrical surfaces, i.e., at $r = r_{1}$ and $r = r_{2}$, results from missing kernel support during post-processing. This phenomenon likewise occurs for both methods, but at a varying degree. Finally, the particle distribution at time $t = 2.0$ is shown in Figure~\ref{fig:example_taylorcouette_particle_distribution} comparing the results of the sliding boundary particle approach with the rigid wall boundary condition~\cite{Adami2012}. In contrast to the rigid wall boundary condition with fixed boundary particles approximating the cylindrical shape, the sliding boundary particle approach does not suffer from geometry discretization errors thus resulting in an improved preservation of the solution symmetry (cf. Figure~\ref{fig:example_taylorcouette_particle_distribution}) and a decreased deviation from the analytical velocity profile (cf. Figure~\ref{fig:example_taylorcouette_angular_vel}). On the other hand, approaches were the cylindrical shape is discretized by boundary particles in ring-shaped arrangement suffer from a disturbed support of the smoothing kernel of fluid particles close to the cylindrical surface, similar than in Figure~\ref{fig:motivation_deforming}.

% begin figure
% end figure

%%
\subsubsection{Laminar flow around a rigid cylinder} \label{subsec:numex_lamflowaroundcyl}

A prominent CFD benchmark problem was proposed by Sch\"afer and Turek et al.~\cite{Schaefer1996} in the year 1996 and since then was considered in a huge variety of publications. The benchmark is concerned with the laminar flow around a rigid cylinder in a channel. Within this publication, the problem is utilized to validate the momentum exchange at the fluid-structure interface, i.e., at the surface of the cylinder, examining characteristic quantities such as the drag and the lift coefficient or the cycle duration of the time-periodic solution. In the following the focus is set on the two-dimensional, unsteady test case 2D-2~\cite{Schaefer1996}.

% begin figure
\begin{figure}[htbp]
\centering
\newcommand*{\scaletext}{1.0}
\newcommand*{\scalefig}{0.5}
\psfrag{L}{\scalebox{\scaletext}{$L$}}
\psfrag{H}{\scalebox{\scaletext}{$H$}}
\psfrag{D}{\scalebox{\scaletext}{$D$}}
\psfrag{cx}{\scalebox{\scaletext}{$x$}}
\psfrag{cy}{\scalebox{\scaletext}{$y$}}
\psfrag{Uin}{\scalebox{\scaletext}{$\vectorbold{u}_{in}$}}
\psfrag{pout}{\scalebox{\scaletext}{$p_{out}$}}
\psfrag{nosl}{\scalebox{\scaletext}{no-slip b.c.}}
\includegraphics[scale=\scalefig]{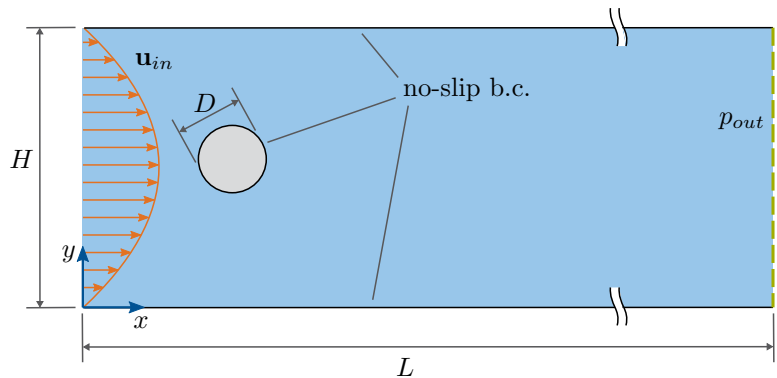}
\caption{Laminar flow around a rigid cylinder: geometry and boundary conditions of the benchmark problem as proposed by Sch\"afer and Turek et al.~\cite{Schaefer1996}.}
\label{fig:example_flowaroundcyl}
\end{figure}
% end figure

Consider a rigid cylinder of diameter~$D=0.1$ with center fixed at position~$\qty(0.2,0.2)$ in a rectangular channel of length~$L=2.2$ and height~$H=0.41$, as illustrated in Figure~\ref{fig:example_flowaroundcyl}. The channel is filled by a Newtonian fluid initially at rest with density~$\rho^{f} = 1.0$ and kinematic viscosity~$\nu^{f} = 1.0 \times 10^{-3}$. It shall be noted that the problem setup is designed intentionally non-symmetric in order to initiate unsteady vortex shedding behind the cylinder. No-slip boundary conditions are applied at the bottom and top channel wall and on the surface of the cylinder. At the channel inflow, a parabolic, time dependent velocity profile~$\vectorbold{u}_{in} = \vectorbold{u}\qty(x=0,y,t)$ is prescribed with components
\begin{equation}
u_{x}\qty(x=0,y,t) = u_{max} \frac{4 y \qty(H - y)}{H^{2}} \, \tau\qty(t) \qand u_{y}\qty(x=0,y,t) = 0.0
\end{equation}
where
\begin{equation}
\tau\qty(t) = \begin{cases}
  \frac{1}{2} \qty(1-\cos{\qty(\frac{\pi}{2}t)}) & \qif t < 2.0 \\
  1.0 & \qotherwise
\end{cases}
\, .
\end{equation}
The maximum inflow velocity is set to~$u_{max} = 1.5$ resulting in a Reynolds number $Re = \flatfrac{u_{mean}D}{\nu^{f}} = 100$ with mean velocity $u_{mean} = \flatfrac{2}{3} u_{max} = 1.0$ for all times~$t \geq 2.0$. At the channel outflow a zero pressure condition $p_{out} = p\qty(x=L,y,t) = 0.0$ is applied.

The fluid domain is discretized by fluid particles with initial particle spacing $\Delta{}x = 2.0 \times 10^{-3}$. The smoothing length~$h$ is set equal to the initial particle spacing~$\Delta{}x$ resulting in a support radius~$r_{c} = 6.0 \times 10^{-3}$ of the smoothing kernel. An artificial speed of sound~$c = 12.5$ is chosen for the fluid phase leading to a reference pressure~$p_{0} = 156.25$. The background pressure is set to~$p_{b} = 312.5$ and is on the order of the reference pressure as proposed by~\cite{Adami2013}. The bottom and top channel walls are modeled utilizing boundary particles according to~\cite{Adami2012} with spacing equal to the initial particle spacing~$\Delta{}x$. On account of the fact that the cylinder is fixed and undeformable only the surface of the cylinder is regularly discretized by 48 surface elements of same size that are considered in the computation of the fluid field, i.e., the structural field is not solved. The unsteady flow simulation is solved for times~$t \in \qty[0, 8.0]$ with a time step size of~$\Delta{}t = 4.0 \times 10^{-5}$ based on the time step size conditions defined in equation~\eqref{eq:sph_timestepcond}.

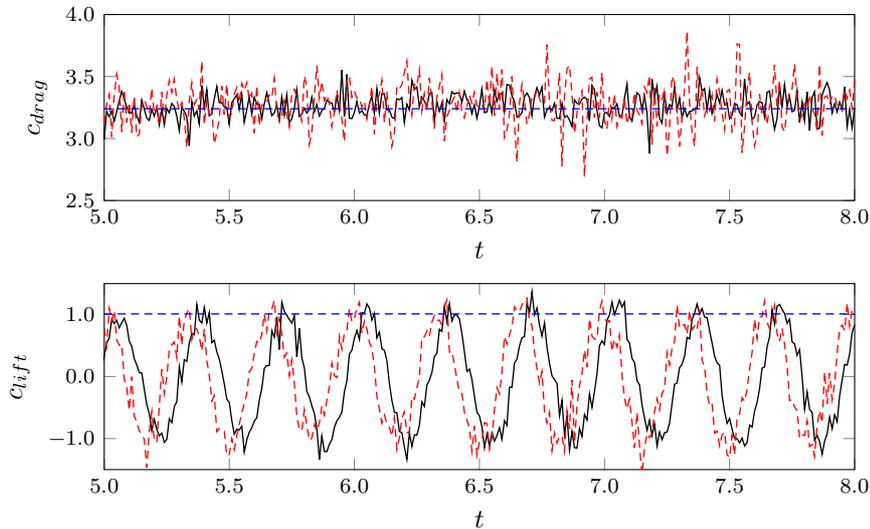
\begin{figure}[htbp]
\centering
% begin subfigure
\subfigure %[caption of subfigure]
{
\begin{tikzpicture}[trim axis left,trim axis right]
\begin{axis}
[
width=0.6\textwidth,
height=0.15\textwidth,
xmin=5.0, xmax=8.0,
ymin=2.5, ymax=4.0,
xtick={5.0, 5.5, 6.0, 6.5, 7.0, 7.5, 8.0},
xticklabel style={/pgf/number format/precision=1, /pgf/number format/fixed, /pgf/number format/fixed zerofill},
yticklabel style={/pgf/number format/precision=1, /pgf/number format/fixed, /pgf/number format/fixed zerofill},
xlabel={$t$},
ylabel={$c_{drag}$},
]
\addplot [color=black,solid,line width=0.5pt] table [x expr={(\thisrow{"Time"})}, y expr={(\thisrow{"drag_coeff (stats)"})}, col sep=comma] {data/flowaroundcyl_drag_lift_coeff.csv};
\addplot [color=red,densely dashed,line width=0.5pt] table [x expr={(\thisrow{"Time"})}, y expr={(\thisrow{"drag_lift (0) (stats)"})}, col sep=comma] {data/flowaroundcyl_drag_lift_coeff_bdry.csv};
%
%\addplot [color=blue,densely dashed,line width=0.5pt] coordinates{(5.0,3.22) (8.0,3.22)};
\addplot [color=blue,densely dashed,line width=0.5pt] coordinates{(5.0,3.24) (8.0,3.24)};
\end{axis}
\end{tikzpicture}
}
% end subfigure
% begin subfigure
\subfigure %[caption of subfigure]
{
\begin{tikzpicture}[trim axis left,trim axis right]
\begin{axis}
[
width=0.6\textwidth,
height=0.15\textwidth,
xmin=5.0, xmax=8.0,
ymin=-1.5, ymax=1.5,
xtick={5.0, 5.5, 6.0, 6.5, 7.0, 7.5, 8.0},
xticklabel style={/pgf/number format/precision=1, /pgf/number format/fixed, /pgf/number format/fixed zerofill},
yticklabel style={/pgf/number format/precision=1, /pgf/number format/fixed, /pgf/number format/fixed zerofill},
xlabel={$t$},
ylabel={$c_{lift}$},
]
\addplot [color=black,solid,line width=0.5pt] table [x expr={(\thisrow{"Time"})}, y expr={(\thisrow{"lift_coeff (stats)"})}, col sep=comma] {data/flowaroundcyl_drag_lift_coeff.csv};
\addplot [color=red,densely dashed,line width=0.5pt] table [x expr={(\thisrow{"Time"})}, y expr={(\thisrow{"drag_lift (1) (stats)"})}, col sep=comma] {data/flowaroundcyl_drag_lift_coeff_bdry.csv};
%
%\addplot [color=blue,densely dashed,line width=0.5pt] coordinates{(5.0,0.99) (8.0,0.99)};
\addplot [color=blue,densely dashed,line width=0.5pt] coordinates{(5.0,1.01) (8.0,1.01)};
\end{axis}
\end{tikzpicture}
}
% end subfigure
\caption{Laminar flow around a rigid cylinder: drag coefficient~$c_{drag}$ and lift coefficient~$c_{lift}$ using the proposed sliding boundary particle approach (black solid line) and the rigid wall boundary condition~\cite{Adami2012} (red dashed line) compared to the upper bounds given in reference solution~\cite{Schaefer1996} (blue dashed line).}
\label{fig:example_flowaroundcyl_drag_lift_coeff}
\end{figure}

\begin{figure}[htbp]
\centering
% begin subfigure
\subfigure [time $t = 6.90$]
{
\includegraphics[width=0.8\textwidth]{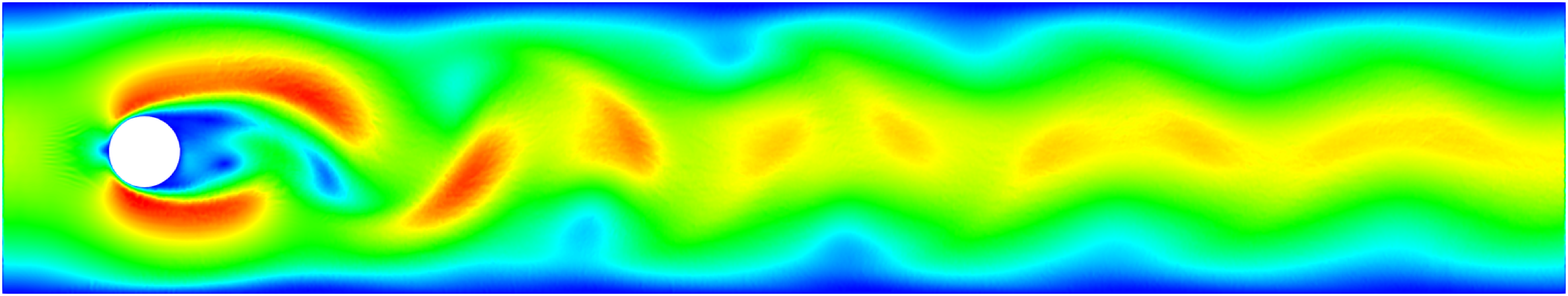}
}
% end subfigure
% begin subfigure
\subfigure [time $t = 6.98$]
{
\includegraphics[width=0.8\textwidth]{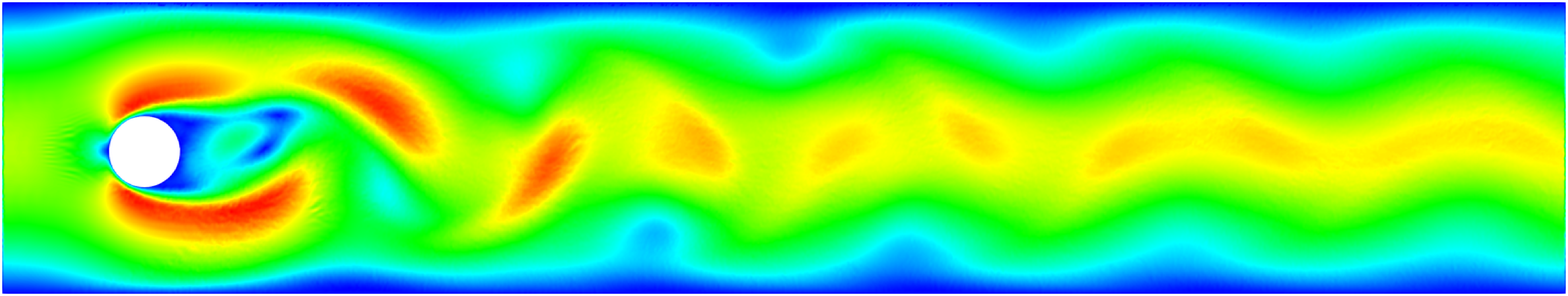}
}
% end subfigure
% begin subfigure
\subfigure [time $t = 7.06$]
{
\includegraphics[width=0.8\textwidth]{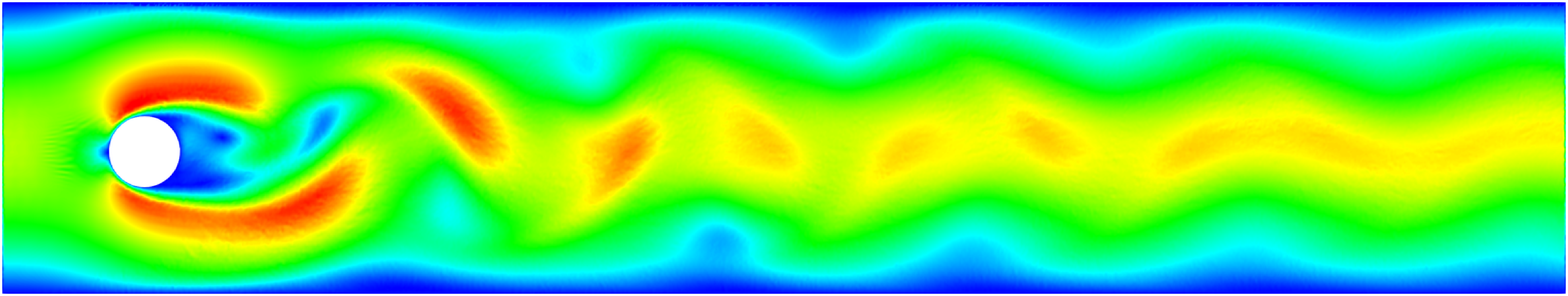}
}
% end subfigure
% begin subfigure
\subfigure [time $t = 7.14$]
{
\includegraphics[width=0.8\textwidth]{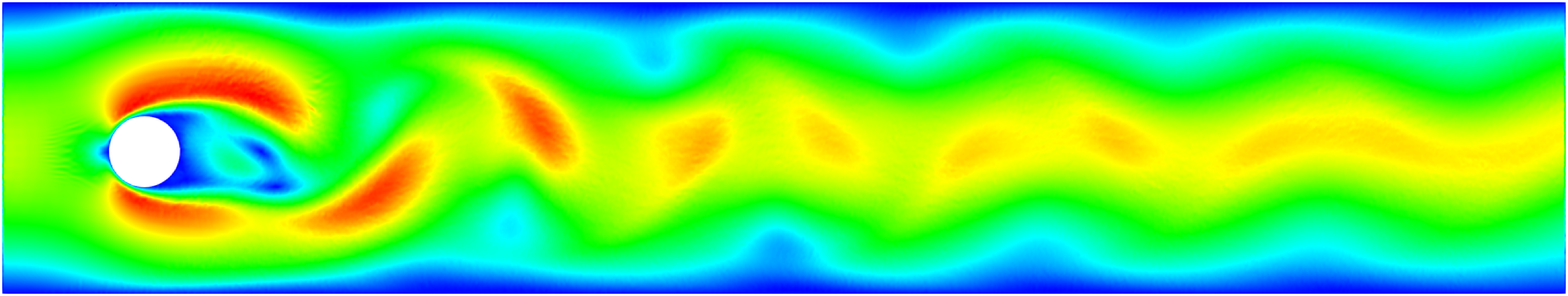}
}
% end subfigure
\caption{Laminar flow around a rigid cylinder: magnitude of the fluid velocity field ranging from $0.0$ (blue) to $2.2$ (red) for a periodic cycle from $t_{0} = 6.90$ to $t_{1} = 7.23$ at four equidistant points in time.}
\label{fig:example_flowaroundcyl_velocity_field}
\end{figure}

To allow for a quantitative comparison of the obtained results with existing reference solutions, the drag and the lift coefficient are defined as
\begin{equation}
c_{drag} = \frac{2 f_{drag}}{\rho^{f} u_{mean}^{2} D}
\qand
c_{lift} = \frac{2 f_{lift}}{\rho^{f} u_{mean}^{2} D}
\end{equation}
where~$f_{drag}$ and~$f_{lift}$ denote the forces in $x$- respectively $y$-direction acting on the cylinder obtained from the sum of all force contributions of fluid particles acting on interface elements of the discretized surface of the cylinder, cf. equation~\eqref{eq:inter_resforce}. Figure~\ref{fig:example_flowaroundcyl_drag_lift_coeff} shows the drag coefficient~$c_{drag}$ and the lift coefficient~$c_{lift}$ obtained for the fully developed time-periodic solution after approximately~$t=5.0$. Both drag and lift coefficient show typical fluctuations as common in SPH-based simulations (similar to an example in~\cite{Adami2013}), that result from disturbances of the density field~\cite{Morris1997} due to relative particle movement. Besides that, the obtained results are in good agreement to the lower bound ($c_{drag} = 3.2200$, $c_{lift} = 0.9900$) and upper bound ($c_{drag} = 3.2400$, $c_{lift} = 1.0100$) of the maximum drag and lift coefficient given by~\cite{Schaefer1996}. The shape of the curve of lift coefficient~$c_{lift}$ allows identifying periodic cycles of the solution with approximate cycle duration~$t_{cycle} = 0.33$ in close agreement to the result of~\cite{TurekOnlineCFD}. In addition, the example is computed discretizing the cylinder with fixed boundary particles based on an implementation of the rigid wall boundary condition~\cite{Adami2012}. The results in form of the drag coefficient~$c_{drag}$ and the lift coefficient~$c_{lift}$ are compared to those obtained with the proposed sliding boundary particle approach, cf. Figure~\ref{fig:example_flowaroundcyl_drag_lift_coeff}, and likewise show the observed typical fluctuations. Note that the visible phase shift in the time-periodic solution of the lift coefficient~$c_{lift}$ is stemming from roundoff errors that influence the initiation of vortex shedding. Finally, Figure~\ref{fig:example_flowaroundcyl_velocity_field} shows the magnitude of the fluid velocity field for a periodic cycle from $t_{0} = 6.90$ to $t_{1} = 7.23$ at four equidistant points in time. At time~$t = 6.98$ the present results of the velocity field visualized in Figure~\ref{fig:example_flowaroundcyl_velocity_field} are qualitatively in good agreement to the results of~\cite{TurekOnlineCFD}. Altogether, the results of the CFD benchmark problem obtained with the sliding boundary particle approach represent the given reference solutions~\cite{Schaefer1996,TurekOnlineCFD} both quantitatively and qualitatively in good approximation and further showcase the capabilities of the novel formulation to accurately model the momentum exchange at the fluid-structure interface.

% begin figure
% end figure

% begin figure
% end figure

%%
\subsubsection{Isochoric deformation of a box filled with a fluid} \label{subsec:numex_deforming_box}

This example aims to demonstrate the advantages of the proposed sliding boundary particle approach over fixed (material) boundary particle methods in the case of large deformations at the fluid-structure interface. To this end, an academic example is examined utilizing both the proposed sliding boundary particle approach and an implementation of the rigid wall boundary condition~\cite{Adami2012} with boundary particles fixed to material points of the structure.

An initially quadratic structural box with inner edge length~$b = 0.1$ and wall thickness~$d = 0.015$ is filled by a Newtonian fluid initially at rest with density~$\rho^{f} = 1.0$ and kinematic viscosity~$\nu^{f} = 1.0 \times 10^{-2}$. An isochoric deformation of the structural box to obtain a rectangular shape (with final edge lengths~$b_{x} = 0.2$ and~$b_{y} = 0.05$ starting from $t = 2.5$) is prescribed, defined by the deformation gradient
\begin{equation}
\vectorbold{F} = \mqty[ \lambda\qty(t) & 0 \\ 0 & \flatfrac{1.0}{\lambda\qty(t)} ]
\qq{where}
\lambda\qty(t) = \begin{cases}
  1.0 + 0.4 t & \qif t < 2.5 \\
  2.0 & \qotherwise
\end{cases}
\, ,
\end{equation}
and accordingly with $\det\vectorbold{F} = 1.0$. It follows, that also the volume of the fluid within the structural box remains constant at all times. Consequently, in the final static equilibrium state the fluid density is expected to be constant throughout the entire fluid domain.

The fluid domain within the structural box is discretized by fluid particles with initial particle spacing~$\Delta{}x = 5.0 \times 10^{-3}$. The smoothing length~$h$ of the smoothing kernel is set equal to the initial particle spacing~$\Delta{}x$ resulting in a support radius~$r_{c} = 1.5 \times 10^{-2}$. For the fluid phase, an artificial speed of sound~$c = 1.0$ is chosen, leading to a reference pressure~$p_{0} = 1.0$. The background pressure~$p_{b}$ is set equal to the reference pressure~$p_{0}$. The walls of the structural box are either modeled by surface elements when using the proposed sliding boundary particle approach or by boundary particles fixed to material points of the structure when using the rigid wall boundary condition~\cite{Adami2012}. The problem is solved for times~$t \in \qty[0, 10.0]$ with time step size~$\Delta{}t = 3.125 \times 10^{-4}$, cf. equation~\eqref{eq:sph_timestepcond}.

\begin{figure}[htbp]
\centering
% begin subfigure
\subfigure [Initial configuration with fluid particles (gray) and boundary particles (black).]
{
\includegraphics[width=0.45\textwidth]{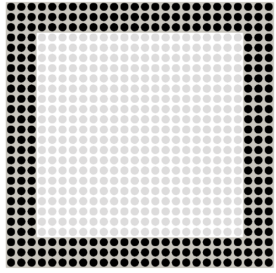}
}
% end subfigure
\hspace{0.05\textwidth}
% begin subfigure
\subfigure [Final configuration with fluid particles colored with fluid density ranging from 0.9 (blue) to 1.1 (red) and boundary particles (black).]
{\includegraphics[width=0.45\textwidth]{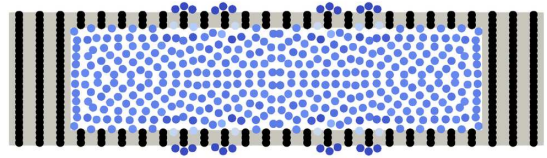}
}
% end subfigure
\caption{Isochoric deformation of a box filled with a fluid: particle distribution obtained using the rigid wall boundary condition~\cite{Adami2012} with fixed (material) boundary particles at points in time $t_{0} = 0.0$ and $t_{1} = 10.0$.}
\label{fig:example_deforming_box_bdry}
\end{figure}

\begin{figure}[htbp]
\centering
% begin subfigure
\subfigure [Initial configuration with fluid particles (gray) and surface elements (blue).]
{
\includegraphics[width=0.45\textwidth]{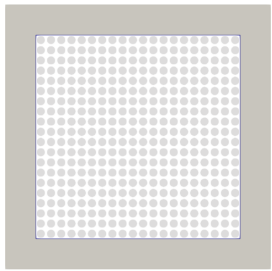}
}
% end subfigure
\hspace{0.05\textwidth}
% begin subfigure
\subfigure [Final configuration with fluid particles colored with fluid density ranging from 0.9 (blue) to 1.1 (red) and an illustration of the virtual boundary particles (black) belonging to two characteristic fluid particles.]
{
\includegraphics[width=0.45\textwidth]{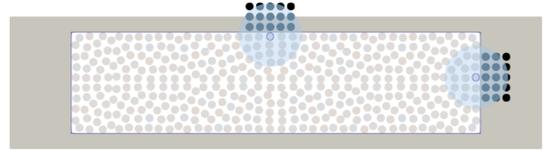}
}
% end subfigure
\caption{Isochoric deformation of a box filled with a fluid: particle distribution obtained using the proposed sliding boundary particle approach at points in time $t_{0} = 0.0$ and $t_{1} = 10.0$.}
\label{fig:example_deforming_box_wall}
\end{figure}

Figure~\ref{fig:example_deforming_box_bdry} shows the particle distribution obtained using the rigid wall boundary condition~\cite{Adami2012} with fixed (material) boundary particles in the initial configuration and at time~$t = 10.0$. Presribing the deformation of the structural box naturally also distorts the initially regular arrangement of boundary particles fixed to material points of the structure, i.e., the boundary particle spacing is streched in horizontal direction and compressed in vertical direction, which is clearly visible at time~$t = 10.0$. As a consequence, the support of the smoothing kernel of a fluid particle close to the interface is disturbed, also influencing the density (and pressure) field in the deformed fluid domain. Eventually, leakage of fluid particles through the fluid-structure interface occurs when the boundary particle spacing becomes too large, and accordingly, the fluid density within the structural box is significantly reduced with an average density error of approximately $7.5 \, \%$. The results obtained with the proposed sliding boundary particle approach are shown in Figure~\ref{fig:example_deforming_box_wall}. For the purposes of illustration, at time~$t = 10.0$ the virtual boundary particles belonging to a fluid particle close to the upper edge and to a fluid particle close to the right edge are shown. Here the full benefits of the proposed sliding boundary particle approach become obvious: full support of the smoothing kernel of a fluid particle close to the interface is retained by a transient set of regularly arranged virtual boundary particles. As a result, an undisturbed density (and pressure) field is achieved in the deformed fluid domain, and consequently, no leakage of fluid particles through the fluid-structure interface occurs. Altogether, this example illustrates the advantages of the proposed sliding boundary particle approach over fixed (material) boundary particle methods when considering large deformations of the fluid-structure interface.

% begin figure
% end figure

% begin figure
% end figure

%%
\subsection{Validation of the fluid-structure interaction framework}

Additional complexity is added by considering freely moving and deformable structures stressing the coupling of fluid and structural field following a Dirichlet-Neumann partitioned approach. Consequently, in the following examples also the structural field is solved. Analytical solutions and reference solutions given in the literature are used to validate the obtained results in quantitative and qualitative manner.

\subsubsection{A rigid cylinder floating in a shear flow} \label{subsec:numex_cylindershearflow}

The following example is based on the studies~\cite{Feng1994, Feng2002} stating that a rigid cylinder floating in a shear flow in a channel always migrates to the center of the channel independent of its initial position and velocity. Here, this example serves as a further validation of the proposed method considering rigid body motion of the structural field. For validation, the obtained results are compared to~\cite{Hashemi2012} where both the fluid and the solid field are discretized using SPH.

\begin{figure}[htbp]
\centering
\newcommand*{\scaletext}{1.0}
\newcommand*{\scalefig}{0.5}
\psfrag{L}{\scalebox{\scaletext}{$L$}}
\psfrag{H}{\scalebox{\scaletext}{$H$}}
\psfrag{D}{\scalebox{\scaletext}{$D$}}
\psfrag{cx}{\scalebox{\scaletext}{$x$}}
\psfrag{cy}{\scalebox{\scaletext}{$y$}}
\psfrag{Uw}{\scalebox{\scaletext}{$\flatfrac{u_{w}}{2}$}}
\psfrag{pbc}{\scalebox{\scaletext}{periodic boundary}}
\psfrag{nosl}{\scalebox{\scaletext}{no-slip b.c.}}
\includegraphics[scale=\scalefig]{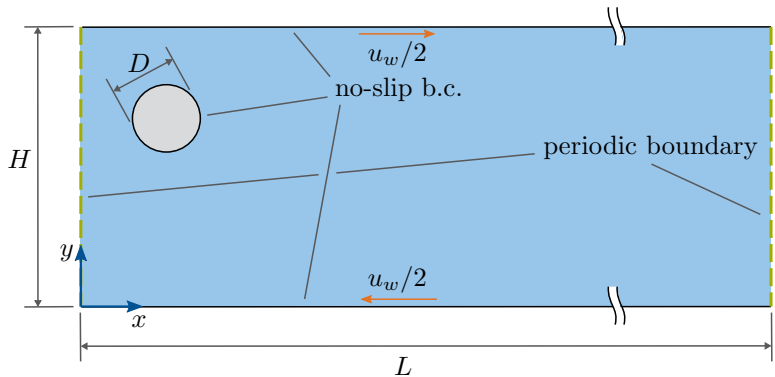}
\caption{A rigid cylinder floating in a shear flow: geometry and boundary conditions of the problem.}
\label{fig:example_cylindershearflow}
\end{figure}

A rigid cylinder of diameter~$D = 0.0025$ allowed to move freely is initially placed at position~$\qty(0.002, 0.0075)$ in a rectangular channel of length~$L=0.1$ and height~$H=0.01$, as illustrated in Figure~\ref{fig:example_cylindershearflow}. The remainder of the channel is occupied by a Newtonian fluid with density~$\rho^{f} = 1.0$ and kinematic viscosity~$\nu^{f} = 5.0 \times 10^{-6}$. The bottom and top channel walls move with a velocity~$\flatfrac{u_{w}}{2} = 0.01$ in opposite direction inducing a shear flow in the channel under consideration of no-slip boundary conditions on all fluid-structure interfaces. The Reynolds number of the problem is $Re = \flatfrac{u_{w} D^{2}}{4 \nu^{f} H} = 0.625$~\cite{Feng2002, Hashemi2012} taking into account the diameter of the cylinder~$D$ and the channel height~$H$. At the left and right end of the channel, periodic boundary conditions are applied.

% begin figure
% end figure

In this example, the fluid domain is discretized by fluid particles with initial particle spacing~$\Delta{}x = 1.0 \times 10^{-4}$. The smoothing length~$h$ is equal to the initial particle spacing~$\Delta{}x$ leading to a support radius~$r_{c} = 3.0 \times 10^{-4}$ of the smoothing kernel. An artificial speed of sound~$c = 0.25$ is chosen, resulting in a reference pressure~$p_{0} = 0.0625$ for the fluid phase. The background pressure~$p_{b}$ is set equal to the reference pressure~$p_{0}$. The motion of the bottom and top channel walls is modeled using moving boundary particles according to~\cite{Adami2012}. A Saint Venant-Kirchhoff model with relatively high Young's modulus~$E^{s} = 1.0 \times 10^{6}$ and Poisson's ratio~$\nu^{s} = 0.4$ is applied for the structure in order to penalize deformation of the cylinder and allow primarily rigid body motions. The cylinder is regularly discretized by 144 first-order elements with 48 surface elements on the surface of the cylinder. Convergence of the iterative coupling algorithm is checked based on the tolerance~$\epsilon = 1.0 \times 10^{-8}$ in equation~\eqref{eq:coup_convergence}. The problem is solved for times~$t \in \qty[0, 60.0]$ with a time step size of~$\Delta{}t = 1.0 \times 10^{-4}$.

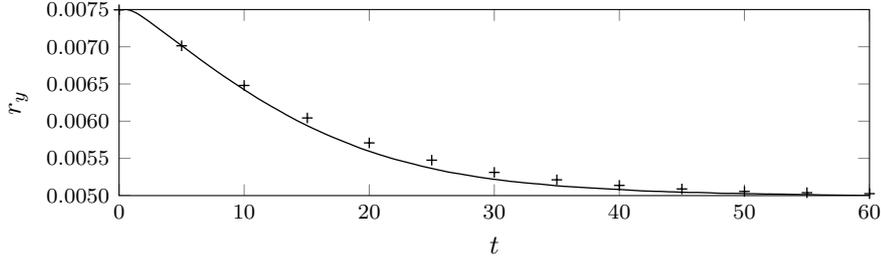
\begin{figure}[htbp]
\centering
\begin{tikzpicture}[trim axis left,trim axis right]
\begin{axis}
[
width=0.6\textwidth,
height=0.15\textwidth,
xmin=0.0, xmax=60.0,
ymin=0.005, ymax=0.0075,
scaled ticks=false,
ytick={0.005, 0.0055, 0.006, 0.0065, 0.007, 0.0075},
yticklabel style={/pgf/number format/precision=4, /pgf/number format/fixed, /pgf/number format/fixed zerofill},
xlabel={$t$},
ylabel={$r_{y}$},
]
\addplot [color=black,solid,line width=0.5pt] table [x expr={(\thisrow{"Time"})}, y expr={0.005+(\thisrow{"Y (stats)"})}, col sep=comma] {data/cylinshearflow_ypos.csv};
\addplot [color=black,only marks,mark=+] table [x expr={(\thisrow{"time"})}, y expr={0.005+(\thisrow{"ypos"})}, col sep=comma] {data/cylinshearflow_hashemi2012_data.csv};
%\addplot [color=red,only marks,mark=o] table [x expr={(\thisrow{"time"})}, y expr={0.005+(\thisrow{"ypos"})}, col sep=comma] {data/cylinshearflow_feng1994_data.csv};
%\addplot [color=blue,only marks,mark=x] table [x expr={(\thisrow{"time"})}, y expr={(\thisrow{"ypos"})}, col sep=comma] {data/cylinshearflow_feng2002_data.csv};
%
\end{axis}
\end{tikzpicture}
\caption{A rigid cylinder floating in a shear flow: vertical position~$r_{y}$ of the center of the cylinder in the channel computed by the method developed in this article (solid line) compared to reference solution~\cite{Hashemi2012} (crosses).}
\label{fig:example_cylindershearflow_vertical_pos}
\end{figure}

The vertical position of the center of the cylinder in the channel over time~$t$ is displayed in Figure~\ref{fig:example_cylindershearflow_vertical_pos}. The cylinder migrates to the center line of the channel as expected. In addition, a quantitative comparison to the results given in~\cite{Hashemi2012} shows good agreement for the dynamcis of the solution.

% begin figure
% end figure

%%
\subsubsection{Flow-induced oscillations of a flexible beam attached to a rigid cylinder} \label{subsec:numex_flexiblebeam}

Based on the benchmark problem of a laminar flow around a rigid cylinder in a channel~\cite{Schaefer1996}, cf. Section~\ref{subsec:numex_lamflowaroundcyl}, a FSI benchmark was proposed by Turek and Hron~\cite{Turek2006} as modification of an example first described in~\cite{Ramm1998}. The purpose of the example is to study flow-induced oscillations of a flexible beam attached to a rigid cylinder in a channel flow. In the following the two-dimensional test case FSI2~\cite{Turek2006} is considered that is characterized by large structural displacements.

\begin{figure}[htbp]
\centering
\newcommand*{\scaletext}{1.0}
\newcommand*{\scalefig}{0.5}
\psfrag{L}{\scalebox{\scaletext}{$L$}}
\psfrag{H}{\scalebox{\scaletext}{$H$}}
\psfrag{bl}{\scalebox{\scaletext}{$l$}}
\psfrag{bh}{\scalebox{\scaletext}{$h$}}
\psfrag{D}{\scalebox{\scaletext}{$D$}}
\psfrag{cx}{\scalebox{\scaletext}{$x$}}
\psfrag{cy}{\scalebox{\scaletext}{$y$}}
\psfrag{Uin}{\scalebox{\scaletext}{$\vectorbold{u}_{in}$}}
\psfrag{pout}{\scalebox{\scaletext}{$p_{out}$}}
\psfrag{nosl}{\scalebox{\scaletext}{no-slip b.c.}}
\includegraphics[scale=\scalefig]{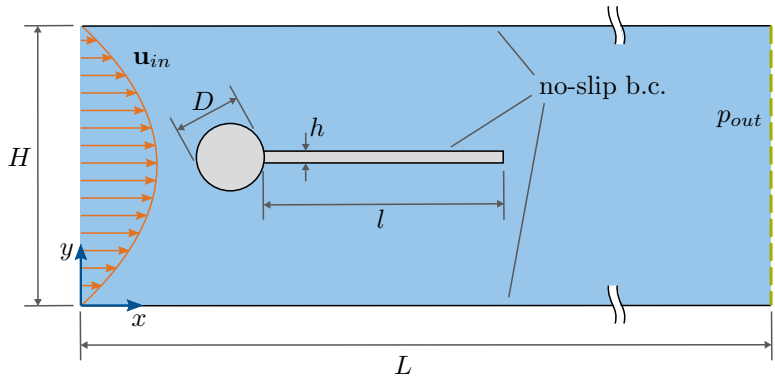}
\caption{Flow-induced oscillations of a flexible beam attached to a rigid cylinder: geometry and boundary conditions of the benchmark problem as proposed by Turek and Hron~\cite{Turek2006}.}
\label{fig:example_flexiblebeam}
\end{figure}

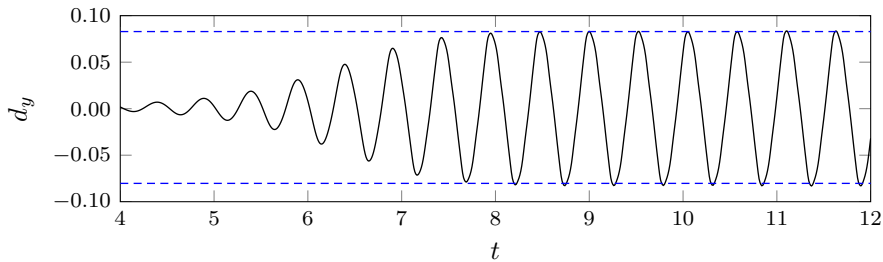
\begin{figure}[htbp]
\centering
\begin{tikzpicture}[trim axis left,trim axis right]
\begin{axis}
[
width=0.6\textwidth,
height=0.15\textwidth,
xmin=4.0, xmax=12.0,
ymin=-0.1, ymax=0.1,
scaled ticks=false,
yticklabel style={/pgf/number format/precision=2, /pgf/number format/fixed, /pgf/number format/fixed zerofill},
xlabel={$t$},
ylabel={$d_{y}$},
]
\addplot [color=black,solid,line width=0.5pt] table [x expr={(\thisrow{"Time"})}, y expr={(\thisrow{"avg displacement (1) (stats)"})}, col sep=comma] {data/flagbehindcyl_tip_disp.csv};
%
%\addplot [color=blue,densely dashed,line width=0.5pt] coordinates{(4.0,-0.07937) (12.0,-0.07937)};
%\addplot [color=blue,densely dashed,line width=0.5pt] coordinates{(4.0,0.08183) (12.0,0.08183)};
%
\addplot [color=blue,densely dashed,line width=0.5pt] coordinates{(4.0,0.0829) (12.0,0.0829)};
\addplot [color=blue,densely dashed,line width=0.5pt] coordinates{(4.0,-0.0803) (12.0,-0.0803)};
\end{axis}
\end{tikzpicture}
\caption{Flow-induced oscillations of a flexible beam attached to a rigid cylinder: vertical displacement~$d_{y}$ of the control point at the tip of the flexible beam using the proposed sliding boundary particle approach (black solid line) compared to the minimum and maximum displacements given in reference solution~\cite{TurekOnlineFSI} (blue dashed line).}
\label{fig:example_flagbehindcyl_tip_disp}
\end{figure}

The problem setup (rectangular channel of length~$L=2.2$ and height~$H=0.41$, rigid cylinder of diameter~$D=0.1$ with center fixed at position~$\qty(0.2,0.2)$) is very equal to the example of a laminar flow around a rigid cylinder discussed in Section~\ref{subsec:numex_lamflowaroundcyl}. In addition, in this example a flexible beam of length~$l = 0.35$ and height~$h = 0.02$ is attached at the downstream end of the rigid cylinder, i.e., at position~$\qty(0.25,0.2)$ as illustrated in Figure~\ref{fig:example_flexiblebeam}. Note that also the length of the channel remains equal to the example in Section~\ref{subsec:numex_lamflowaroundcyl} and is thus slightly shorter than originally proposed for the benchmark~\cite{Turek2006}. A control point needed for evaluation of the results, e.g., in Figure~\ref{fig:example_flagbehindcyl_tip_disp}, is placed at the tip of the flexible beam, i.e., at position~$\qty(0.6,0.2)$ in the undeformed configuration. The fluid properties (Newtonian fluid, density~$\rho^{f} = 1.0$, kinematic viscosity~$\nu^{f} = 1.0 \times 10^{-3}$) remain unchanged compared to the previous example. The density of the flexible structure is set to~$\rho^{s}_{0} = 10.0$ resulting in a density ratio of~$\flatfrac{\rho^{s}_{0}}{\rho^{f}} = 10.0$. A Saint Venant-Kirchhoff model with Young's modulus $E^{s} = 1.4 \times 10^{3}$ (resp. shear modulus~$\mu^{s} = 0.5 \times 10^{3}$) and Poisson's ratio~$\nu^{s} = 0.4$ is utilized to describe the constitutive behavior of the flexible beam. The same boundary conditions as in the example in Section~\ref{subsec:numex_lamflowaroundcyl} (no-slip boundary conditions on all surfaces including the flexible beam, parabolic and time dependent velocity profile at channel inflow with mean velocity~$u_{mean} = 1.0$ for all times~$t \geq 2.0$, zero pressure conditions at channel outflow) are prescribed. The Reynolds number of this example is given to $Re = \flatfrac{u_{mean}D}{\nu^{f}} = 100$.

% begin figure
% end figure

The fluid domain is discretized with fluid particles similar than in the example in Section~\ref{subsec:numex_lamflowaroundcyl} (initial particle spacing $\Delta{}x = 2.0 \times 10^{-3}$, support radius~$r_{c} = 6.0 \times 10^{-3}$ of the smoothing kernel, artificial speed of sound~$c = 12.5$, reference pressure~$p_{0} = 156.25$). In this example, the background pressure is set to~$p_{b} = 1250.0$. Boundary particles according to~\cite{Adami2012} are utilized to model the bottom and top channel walls. The flexible beam as part of the structural domain is discretized by~$35 \times 3$ first-order elements. The surface of the cylinder exposed to the fluid field, i.e., without considering the part where the flexible beam is attached, is discretized by 20 surface elements. Convergence of the iterative coupling of fluid and structural field is based on the tolerance~$\epsilon = 1.0 \times 10^{-8}$, cf. equation~\eqref{eq:coup_convergence}. The FSI problem is solved for times $t \in \qty[0, 12.0]$ with a time step size of~$\Delta{}t = 4.0 \times 10^{-5}$. In this example, convergence of the partitioned coupling loop, cf. Algorithm~\ref{alg:part_fsi}, is reached after an average number of approximately $5.45$ iterations per time step, when averaging over all time steps of the given problem.

\begin{figure}[htbp]
\centering
% begin subfigure
\subfigure [time $t = 10.32$]
{
\includegraphics[width=0.8\textwidth]{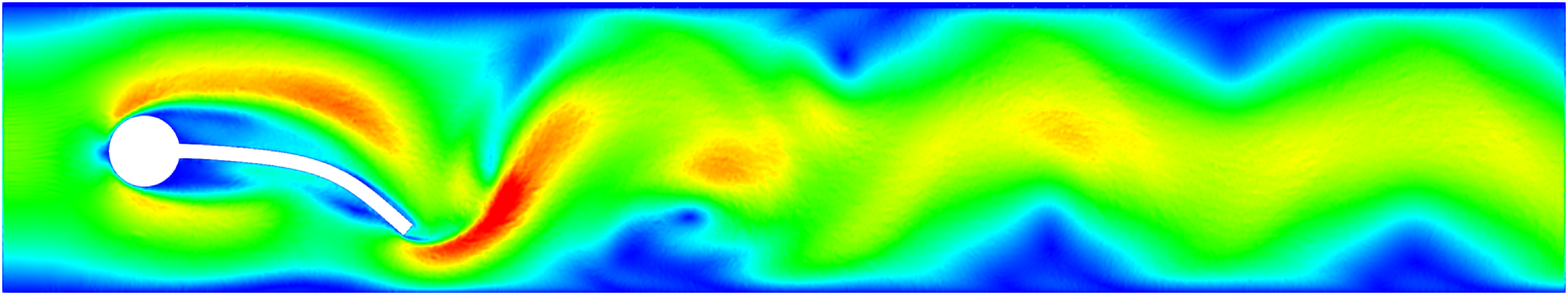}
}
% end subfigure
% begin subfigure
\subfigure [time $t = 10.45$]
{
\includegraphics[width=0.8\textwidth]{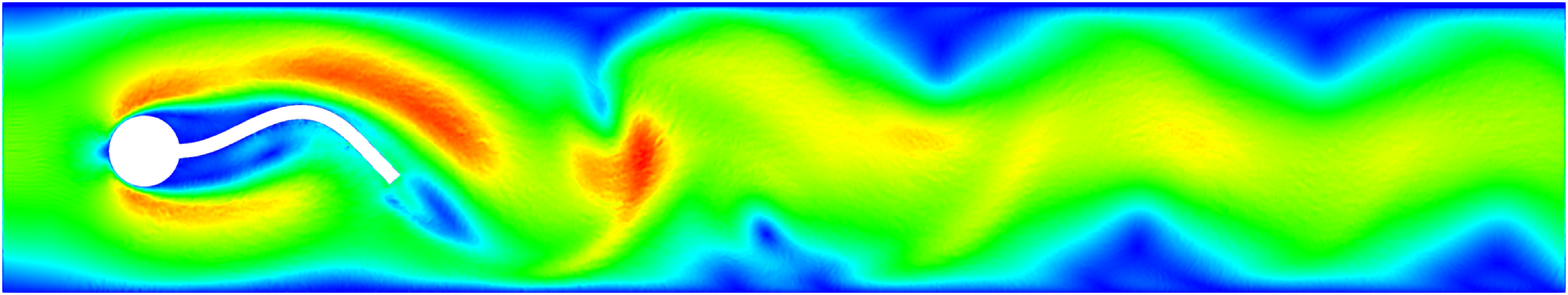}
}
% end subfigure
% begin subfigure
\subfigure [time $t = 10.58$]
{
\includegraphics[width=0.8\textwidth]{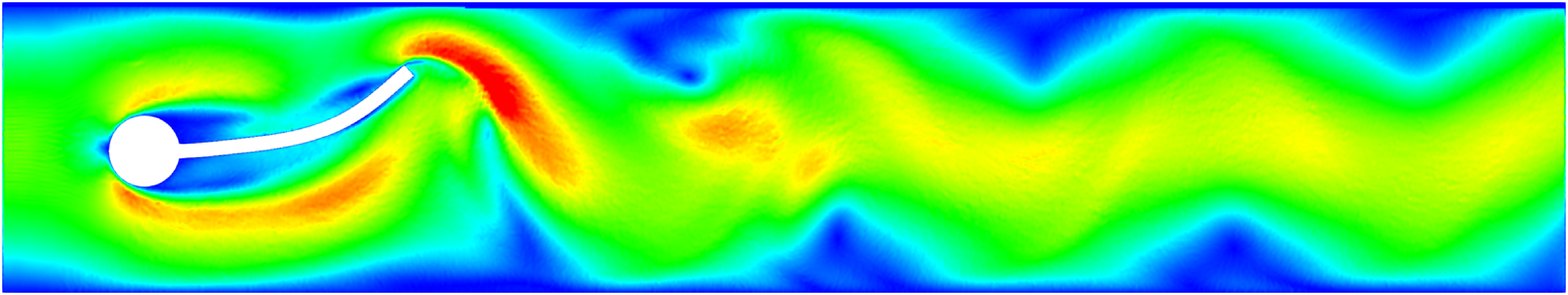}
}
% end subfigure
% begin subfigure
\subfigure [time $t = 10.71$]
{
\includegraphics[width=0.8\textwidth]{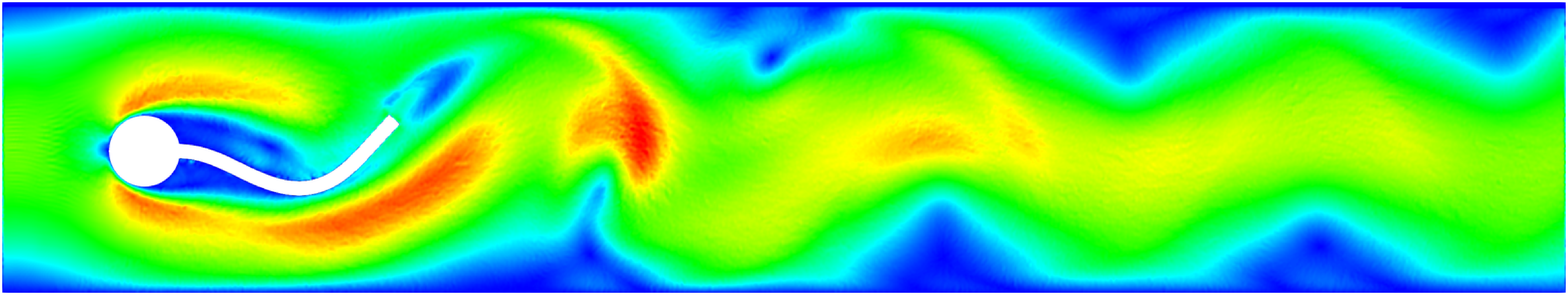}
}
% end subfigure
\caption{Flow-induced oscillations of a flexible beam attached to a rigid cylinder: magnitude of the fluid velocity field ranging from $0.0$ (blue) to $2.5$ (red) and deformation of structure for a periodic cycle from $t_{0} = 10.32$ to $t_{1} = 10.84$ at four equidistant points in time.}
\label{fig:example_flagbehindcyl_velocity_field}
\end{figure}

The vertical displacement~$d_{y}$ of the control point at the tip of the flexible beam is displayed in Figure~\ref{fig:example_flagbehindcyl_tip_disp}. In the present results the minimum and maximum displacement of the control point at the tip of the flexible beam in~$y$-direction are approximately $-0.08269$ and $0.08309$. This is in good agreement with the results given in the literature: \cite{Turek2006} and~\cite{TurekOnlineFSI} report a minimum and maximum displacement of $-0.07937$ and $0.08183$ respectively $-0.0803$ and $0.0829$. The solution of the FSI problem shows time-periodic cycles of the beam deflection after approximately $t = 8.0$ with a cycle duration~$t_{cycle} \approx 0.525$ and a frequency~$f = \flatfrac{1}{t_{cycle}} \approx 1.905$ (averaged over all time-periodic cylces), which is in good agreement with~\cite{Turek2006,TurekOnlineFSI} ($f = 1.90$). In Figure~\ref{fig:example_flagbehindcyl_velocity_field} the magnitude of the fluid velocity field and the deformation of the structure for a periodic cycle from $t_{0} = 10.32$ to $t_{1} = 10.84$ at four equidistant points in time are shown. Especially, at times $t = 10.45$ and $t = 10.71$ the flexible beam experiences strong curvature. This is were approaches discretizing the structural domain by boundary particles fixed to structural material points suffer from a disturbed support of the smoothing kernel of neighboring fluid particles, cf. Figure~\ref{fig:motivation_deforming}. The novel sliding boundary particle approach by definition is not prone to that issue. In summary, the results of the FSI benchmark problem obtained with the sliding boundary particle approach are both quantitatively and qualitatively in good agreement with the given reference solutions~\cite{Turek2006,TurekOnlineFSI}.

% begin figure
% end figure

% begin figure
% end figure

%%
\subsubsection{Inflation of an academic balloon-like problem} \label{subsec:numex_academic_balloon}

The filling process of a highly flexible thin-walled balloon-like container undergoing large deformations is studied in this example, representing a model problem close to potential application scenarios of the proposed scheme in the field of biomechanics.

% begin figure
\begin{figure}[htbp]
\centering
% begin subfigure
\subfigure [Geometry and boundary conditions of the problem based on~\cite{Kuttler2006}.]
{
\newcommand*{\scaletext}{1.0}
\newcommand*{\scalefig}{0.5}
\psfrag{B}{\scalebox{\scaletext}{$B$}}
\psfrag{b}{\scalebox{\scaletext}{$b$}}
\psfrag{d}{\scalebox{\scaletext}{$d$}}
\psfrag{cx}{\scalebox{\scaletext}{$x$}}
\psfrag{cy}{\scalebox{\scaletext}{$y$}}
\psfrag{cz}{\scalebox{\scaletext}{$z$}}
\psfrag{Uin}{\scalebox{\scaletext}{$\vectorbold{u}_{in}$}}
\psfrag{nosl}{\scalebox{\scaletext}{no-slip b.c.}}
\psfrag{str}{\scalebox{\scaletext}{fixed}}
\includegraphics[scale=\scalefig]{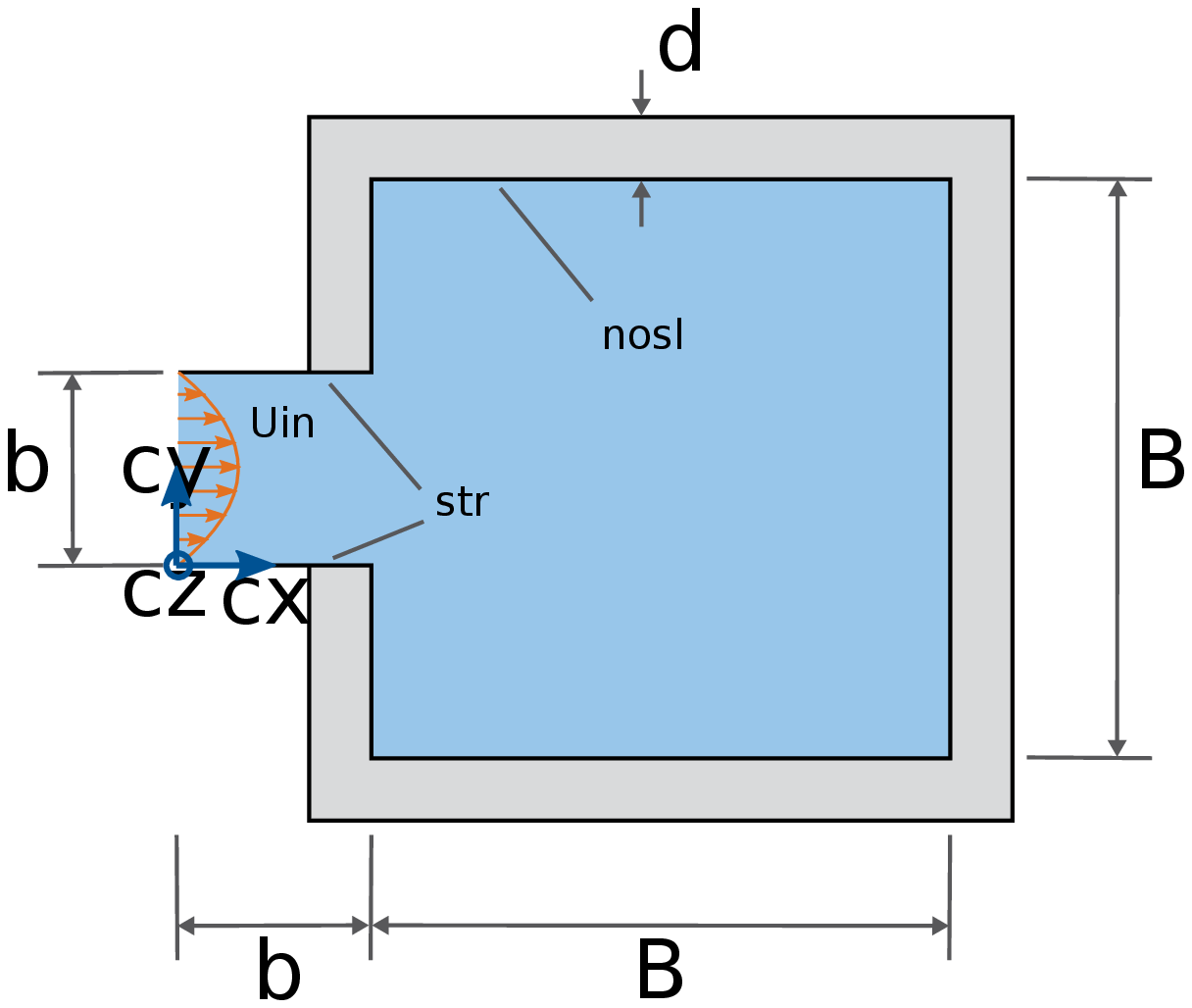}
\label{fig:example_academicballoon_geometry}
}
% end subfigure
\hspace{0.05\textwidth}
% begin subfigure
\subfigure [Volume inside the academic balloon (solid line) compared to analytical solution (crosses).]
{
\hspace{0.05\textwidth}
\begin{tikzpicture}[trim axis left,trim axis right]
\begin{axis}
[
width=0.4\textwidth,
height=0.2\textwidth,
xmin=0.0, xmax=14.0,
ymin=25.0, ymax=60.0,
scaled ticks=false,
xlabel={$t$},
ylabel={$V$},
]
\addplot [color=black,only marks,mark=+] coordinates {(0,27)};
\addplot [color=black,only marks,mark=+,domain=1.0:14.0,samples=14] {27.0+(4.0/9.0)*5.0*(x-0.5)};
\addplot [color=black,solid,line width=0.5pt] table [x expr={(\thisrow{"Time"})}, y expr={(\thisrow{"avg totalvolume (stats)"})}, col sep=comma] {data/academicballoon_volume.csv};
\end{axis}
\end{tikzpicture}
\hspace{0.02\textwidth}
\label{fig:example_academicballoon_volume}
}
% end subfigure
\caption{Inflation of an academic balloon-like problem: setup of the problem and numerical results.}
\label{fig:example_academicballoon}
\end{figure}
% end figure

An initially cubical structural geometry with inner edge length~$B = 3.0$ and wall thickness~$d = 0.2$ is inflated via a quadratic inlet of width and length~$b= 1.0$ by a Newtonian fluid that is initially at rest with density~$\rho^{f} = 1.0$ and kinematic viscosity~$\nu^{f} = 5.0 \times 10^{-1}$, cf. Figure~\ref{fig:example_academicballoon_geometry}. The constitutive behavior of the structure with density $\rho^{s}_{0} = 1.0$ is described by a Saint Venant-Kirchhoff model with Young's modulus $E^{s} = 1.0 \times 10^{2}$ and Poisson's ratio~$\nu^{s} = 0.45$. A similar problem was first proposed in~\cite{Kuttler2006} with the purpose to study and solve the incompressibility dilemma in partitioned fluid-structure interaction with pure dirichlet fluid domains. This dilemma does not exist in our approach, given that SPH uses a weakly compressible approach. Here, the example is recapitulated on a three-dimensional geometry with modified fluid and structural material parameters. No-slip boundary conditions are applied at all fluid-structure interfaces. At the inflow of the balloon-like problem, a parabolic, time dependent velocity profile $\vectorbold{u}_{in} = \vectorbold{u}\qty(x=0,y,z,t)$ with components
\begin{equation}
u_{x}\qty(x=0,y,z,t) = u_{max} \frac{4 y \qty(b - y)}{b^{2}} \frac{4 z \qty(b - z)}{b^{2}} \, \tau\qty(t) \qand u_{y}\qty(x=0,y,z,t) = u_{z}\qty(x=0,y,z,t) = 0.0
\end{equation}
and maximum inflow velocity~$u_{max} = 5.0$ and
\begin{equation}
\tau\qty(t) = \begin{cases}
  \frac{1}{2} \qty(1-\cos{\qty(\pi t)}) & \qif t < 1.0 \\
  1.0 & \qotherwise
\end{cases}
\end{equation}
is prescribed. Note that the origin of the coordinate system $\qty(x,y,z)$ is located at the bottom left corner of the inflow area. Accordingly, the volume inside the academic balloon (without considering the volume of the inlet) can be determined analytically via
\begin{equation} \label{eq:example_academicballoon_analytical}
V\qty(t) = V_{0} + \int_{t} \int_{A_{in}} u_{x} \dd{A} \dd{t}
\end{equation}
for each time $t$ with initial volume $V_{0} = B^{3}$ and inflow area $A_{in} = b^{2}$.

The fluid domain is discretized by fluid particles with initial particle spacing~$\Delta{}x = 4.0 \times 10^{-2}$. The smoothing length $h$ is set equal to the initial particle spacing $\Delta{}x$ resulting in a support radius~$r_{c} = 1.2 \times 10^{-1}$ of the smoothing kernel. The artificial speed of sound is set to $c = 40.0$, hence the reference pressure is $p_{0} = 1600.0$. The background pressure~$p_{b}$ is equal to the reference pressure~$p_{0}$. The walls of the fixed inlet are modeled utilizing boundary particles according to~\cite{Adami2012} with spacing equal to the initial particle spacing~$\Delta{}x$. The balloon-like structural domain is discretized by first-order elements with a cubic shape in the initial configuration and a characteristic element length of $d$ resulting in one element over the wall thickness. This discretization is justified since the focus of this example is set on the coupling of fluid and structural field at the interface rather than a precise prediction of structural quantities such as the deformation field of the tank. The tolerance~$\epsilon = 1.0 \times 10^{-8}$ in equation~\eqref{eq:coup_convergence} is applied for the iterative coupling of fluid and structural field. The FSI Problem is solved for times~$t \in \qty[0, 14.0]$ with a time step size of~$\Delta{}t = 2.5 \times 10^{-4}$. In this example, convergence of the partitioned coupling loop, cf. Algorithm~\ref{alg:part_fsi}, is reached after an average number of approximately $5.88$ iterations per time step, when averaging over all time steps of the given problem.

\begin{figure}[htbp]
\centering
% begin subfigure
\subfigure [time $t = 0.0$]
{
\includegraphics[width=0.3\textwidth]{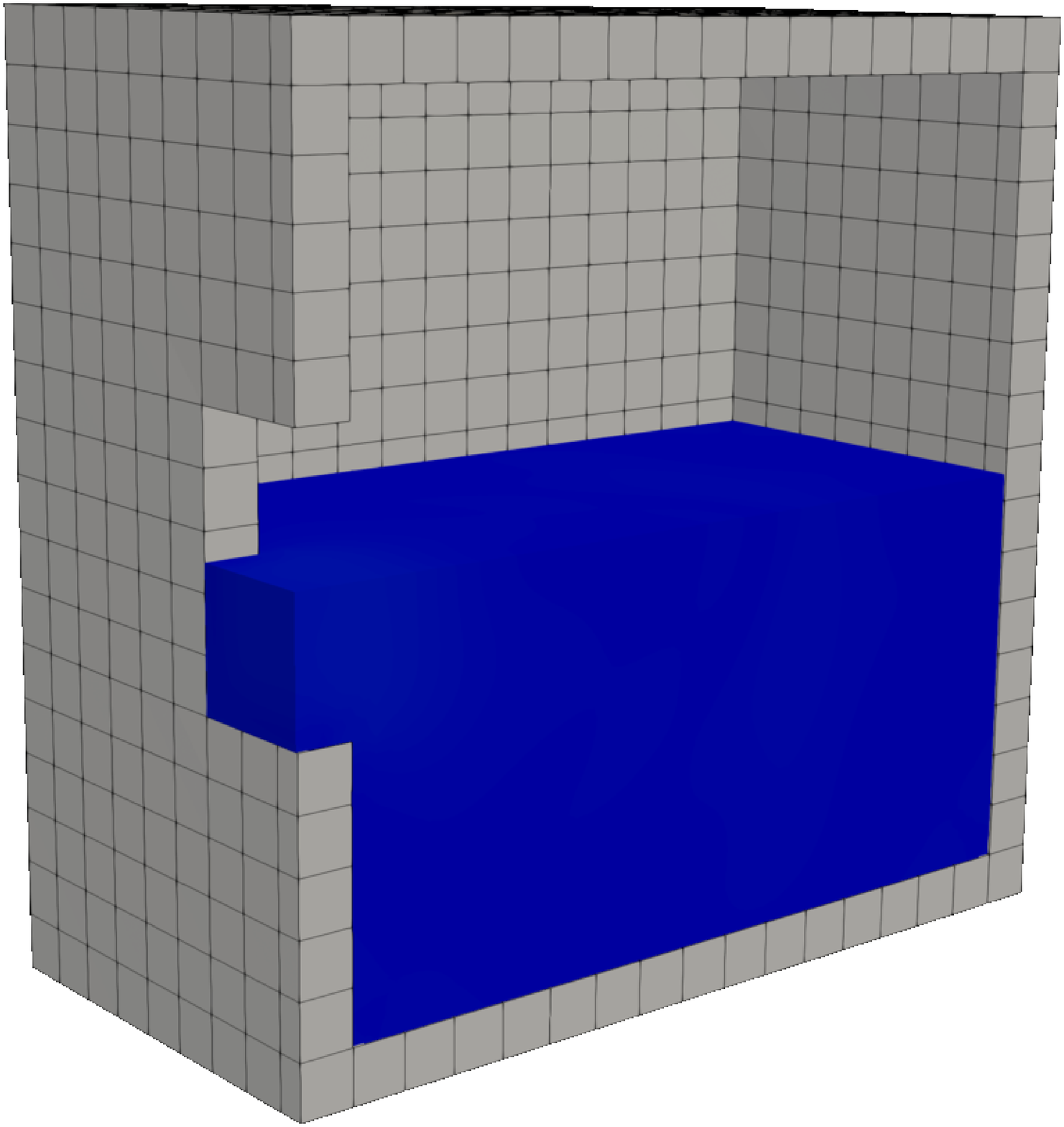}
}
% end subfigure
\hspace{0.1\textwidth}
% begin subfigure
\subfigure [time $t = 12.65$]
{
\includegraphics[width=0.3\textwidth]{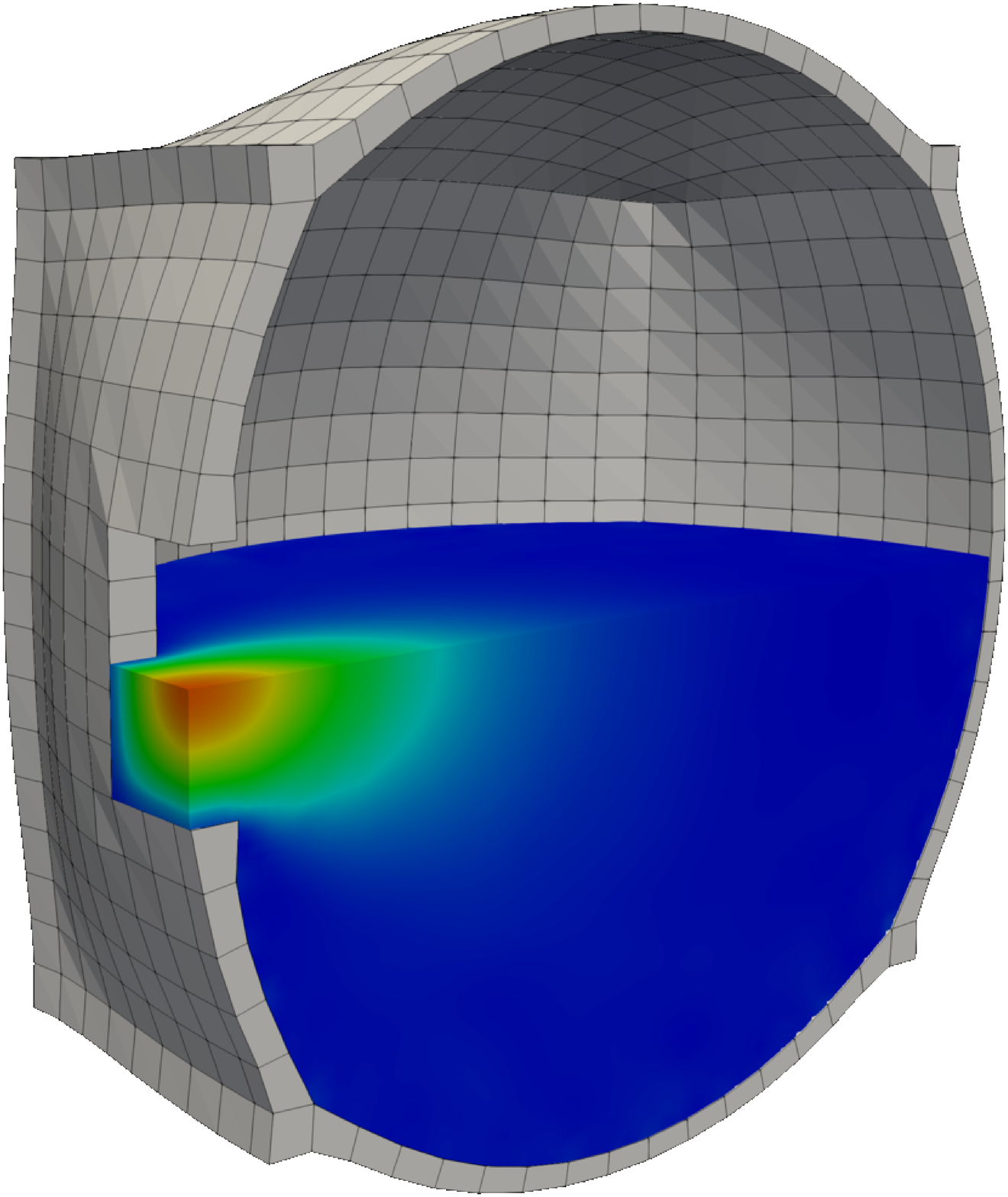}
}
% end subfigure
\caption{Inflation of an academic balloon-like problem: magnitude of the fluid velocity field ranging from $0.0$ (blue) to $5.0$ (red) visualized on a quarter section of the fluid domain and deformation of structure at points in time $t_{0} = 0.0$ and $t_{1} = 12.65$.}
\label{fig:example_academicballoon_deformed}
\end{figure}

The volume inside the academic balloon is determined summing up the effective volumes of respective particles~$j$ following $V = \sum_{j} \flatfrac{m_{j}}{\rho_{j}}$ and compared to the analytical solution~\eqref{eq:example_academicballoon_analytical}, cf. Figure~\ref{fig:example_academicballoon_volume}. The resulting volume is slightly below the analytically determined volume. This can be explained by the weakly compressible approach, cf. Section~\ref{subsec:nummeth_sph_eos}, applied in this SPH formulation leading in this example to a minor compression of the fluid phase with an average density error of approximately $1 \, \%$. Note that conservation of mass and accordingly (within the limits of a weakly compressible approach) conservation of volume is a characteristic property of SPH. By definition, this means that also the number of fluid particles is conserved. Therefore, the obtained results, among others, demonstrate that no leakage of fluid particles through the fluid-structure interface occurs. The rear half of the (inflated) structural geometry and a quarter section of the fluid domain are displayed in Figure~\ref{fig:example_academicballoon_deformed} in the initial state and at time $t = 12.65$. The fluid velocity is post-processed applying SPH approximation~\eqref{eq:sph_postprocessing}. Note that at time $t = 12.65$ the volume inside the academic balloon has doubled, cf. analytical solution~\eqref{eq:example_academicballoon_analytical}. In conclusion, this example is characterized by large structural deformations in form of strong curvature and stretch reaping the full benefits of the proposed sliding boundary particle approach in contrast to fixed (material) boundary particle methods. In the presence of large structural deformations, the latter class of boundary particle methods is characterized by insufficient kernel support, and eventually such methods are prone to leakage of fluid particles.

% begin figure
% end figure

%%
\section{Conclusion and outlook} \label{sec:concl}

A novel smoothed particle hydrodynamics (SPH) and finite element (FE) coupling scheme for fluid-structure interaction, the sliding boundary particle approach, is presented in this publication. The coupled problem is solved via a Dirichlet-Neumann partitioned approach, with the fluid field (discretized via SPH) being the Dirichlet partition and the structural field (discretized via FE) being the Neumann partition. SPH is a mesh-free computational method that simplifies the treatment of both large deformations in the fluid domain as well as complex flow while avoiding additional methodological and computational effort compared to fully mesh-based methods. Introducing the sliding boundary particle approach for the treatment of deformable and strongly curved boundaries of the SPH domain in an accurate, robust, and computationally cheap manner, constitutes an important aspect of the proposed numerical formulation for solving FSI problems.

Several numerical examples showcase the capabilities of the novel numerical formulation. To begin with, the sliding boundary particle approach is validated examining two-dimensional examples driving certain characteristics of the proposed formulation. The numerical results obtained for the examples of a hydrostatic pressure in a fluid between two parallel plates, cf. Section~\ref{subsec:numex_hydrostatic_pressure}, and a planar Taylor-Couette flow, cf. Section~\ref{subsec:numex_taylorcouette}, are in very good agreement with the respective analytical solutions confirming the capability of the proposed method to model linear pressure profiles near the boundary and to account for no-slip boundary conditions at the boundary as required for high accuracy of the fluid velocity field. In a next step, numerical examples involving dynamic effects and large structural deformations are studied confirming the accuracy and robustness of the proposed formulation. This is, among others, demonstrated showing the results of well-known CFD respectively FSI benchmark problems, cf. Sections~\ref{subsec:numex_lamflowaroundcyl} and \ref{subsec:numex_flexiblebeam}, as proposed in~\cite{Schaefer1996,Turek2006}. Altogether, the obtained numerical results are in very good agreement with the results given in the literature. Finally, a three-dimensional, application-focused example is considered examining the filling process of a highly flexible thin-walled container (cf. Section~\ref{subsec:numex_academic_balloon}).

Future work may focus on an asynchronous time stepping scheme, e.g., a sub-cycling scheme of fluid and structural field, cf. Remark~\ref{rmk:coup_asynchronous_time_stepping}. Such an approach would allow to evolve the solution of the sub-fields with different time step sizes each best suitable for the underlying method respectively solver while reducing the overall computational effort. Besides that, the FSI framework may be extended to multiphase flow including the motion of rigid bodies. The framework developed herein will be a valuable tool for detailed studies of biomechanical problems involving complex flow, e.g., the human stomach during digestion~\cite{Brandstaeter2018, Brandstaeter2019}.

\section*{Acknowledgments}

Funded by the Deutsche Forschungsgemeinschaft (DFG, German Research Foundation) - Projektnummer 350481011, Projektnummer 257981274, Projektnummer 386349077. In addition we gratefully thank Georg Hammerl and Niklas Fehn for their preliminary work our particle implementation is based on, Jonas F. Eichinger for discussions concerning parallel programming concepts, Martin Kronbichler for his advice on code efficiency and performance, and Volker Gravemeier for discussions on various aspects in the field of computational fluid dynamics.

\bibliographystyle{elsarticle-num} 
\bibliography{collection.bib}

\end{document}